\documentclass[twocolumn,prb,showpacs,preprintnumbers,amsmath,amssymb]{revtex4}

\usepackage{graphicx}
\usepackage{psfig}
\usepackage{dcolumn}
\usepackage{bm}

\begin{document}

\title{Spin-flop transition in uniaxial antiferromagnets:\\
magnetic phases, reorientation effects, multidomain states}
\author{A.N.\ Bogdanov$^{1,2}$}
\email{a.bogdanov@ifw-dresden.de}
\author{A.V.\ Zhuravlev$^{2}$}
\author{U.K. R\"o\ss ler$^{1}$}
\email{u.roessler@ifw-dresden.de}

\address{$^1$IFW Dresden,
Postfach 270116, D-01171 Dresden, 
Germany}
\address{$^2$Donetsk Institute for 
Physics and Technology,
340114 Donetsk, Ukraine}

\date{\today}

\begin{abstract}
The classical spin-flop is the  
field-driven first-order reorientation transition
in easy-axis antiferromagnets.
A comprehensive phenomenological theory 
of easy-axis antiferromagnets 
displaying spin-flops is developed. 
It is shown how the hierarchy of magnetic coupling 
strengths in these antiferromagnets 
causes a strongly pronounced \textit{two-scale} 
character in their magnetic phase structure.
In contrast to the major part of the magnetic phase diagram, 
these antiferromagnets near the spin-flop region 
are described by an effective model akin to uniaxial ferromagnets.
For a consistent theoretical description both higher-order 
anisotropy contributions and dipolar stray-fields have
to be taken into account near the spin-flop.
In particular, thermodynamically stable multidomain states 
exist in the spin-flop region, owing to the phase
coexistence at this first-order transition.
For this region, equilibrium spin-configurations
and parameters of the multidomain
states are derived as functions
of the external magnetic field.
The components of the magnetic susceptibility tensor
are calculated for homogeneous and multidomain states 
in the vicinity of the spin-flop.
The remarkable anomalies in these 
measurable quantities
provide an efficient method 
to investigate magnetic states and to determine 
materials parameters in bulk and confined antiferromagnets,
as well as in nanoscale synthetic antiferromagnets.
The method is demonstrated for
experimental data on the magnetic 
properties near the spin-flop region
in the orthorhombic layered 
antiferromagnet (C$_2$H$_5$NH$_3$)$_2$CuCl$_4$.
\end{abstract}


\maketitle


\section{Introduction}\label{intro}

\begin{figure}[bht]
\includegraphics[width=8.5cm]{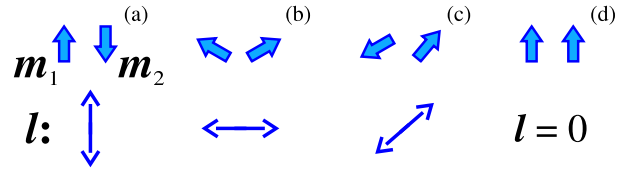}
\caption{
\label{phases}
Basic spin-configurations in collinear uniaxial 
antiferromagnets.
At zero field the magnetization vectors $\mathbf{m}_i$  
are antiparallel (a).
The  magnetic fields  along the easy direction
$H > H_0$ stabilizes the \textit{spin-flop} phase (b).
In magnetic fields deviating from the easy direction 
\textit{canted} states are realized (c).
In a sufficiently strong magnetic fields, $H > H_{\textrm{e}}$,
the spin-flop and canted states transform into 
the saturated \textit{flip} phase (d).
These solutions are degenerate
with respect to the sign of the staggered vector $\mathbf{l}$,
which is represented by two-headed arrows in the bottom panel.
}
\end{figure}

In antiferromagnetic crystals 
with a preferable direction 
of the magnetization,
a sufficiently strong magnetic field
applied along this \textit{easy axis} direction 
``overturns''  the sublattice magnetization vectors
$\mathbf{M}_1$ and  $\mathbf{M}_2$ (Fig.~\ref{phases}).
N\'eel demonstrated this threshold-field effect 
theoretically for classical two-sublattice antiferromagnets 
with sufficiently weak anisotropy in 1936. \cite{Neel36} 
Fifteen years later, this  prediction of 
a jump-like reorientation transition driven 
by an external magnetic field was confirmed 
by experiments 
on the antiferromagnet CuCl$_2 \cdot$2H$_2$O.\cite{Poulis51}
Since that time 
the \textit{spin-flop} transition \cite{Ubbink53}
has been observed 
and was investigated in great detail
for a large group of antiferromagnets
(see, e.g., 
Refs.~$[$\onlinecite{Poulis52,Shapira68,Blazey68,Foner63,Jongh72}$]$
and further  examples and bibliography in
Refs.~$[$\onlinecite{Rives67,UFN88,PRB02}$]$). 
Originally, the name spin-flop (SF) transition
was restricted to the field-driven 
reorientation transition in two-sublattice 
collinear easy-axis antiferromagnets 
following N\'eel's prediction.
However, in many other classes of antiferromagnets
similar mechanisms cause 
different types of field-driven reorientation effects
so that one can speak about a 
class of spin-flop phenomena in a wider sense.
These include field-driven transitions 
in antiferromagnets with a Dzyaloshinskii-Moriya interaction,
\cite{Jongh72,FNT86} 
in multisublattice, \cite{Kastner98,Tsukada01} 
and in noncentrosymmetric antiferromagnets.\cite{PRB02}

The spin-flop transition comprises 
in a simple form main features of magnetic
reorientation and phase transitions
and it gives rise to various
remarkable physical anomalies.
It was found that the spin-flop transition is very
sensitive to direction and values of
the applied field \cite{Rohrer69,FTT82}
and 
specific transitional domain structures are formed
in the spin-flop region.\cite{Wyatt68,UFN88}
These effects have been
observed in MnF$_2$ 
\cite{Dudko72}
and some other antiferromagnets.\cite{RohrerAIP,FNT89}
Theory also  predicts a specific
reconstruction of the domain walls
in the vicinity of the spin-flop
\cite{Jacobs67,UFN88} and 
extraordinary surface effects,
which have been called ``surface spin-flop''.\cite{Mills68}
The investigations on spin-flops
made important contributions 
not only to different research topics in magnetism, 
but also to such general fields of physics
as thermodynamics, \cite{Wong84,UFN88}
nonlinear physics, \cite{Sievers,Fiebig97}
and the theory of phase transitions 
and critical phenomena. \cite{Fisher75}

During the last decade 
spin-flops have been identified and
studied in many novel
bulk antiferromagnets including
such classes 
as antiferromagnetic semiconductors,\cite{Fries97}
organic magnets, \cite{Wernsdorfer02}
in cuprates, such as La$_2$CuO$_4$ and Nd$_2$CuO$_4$, being
base materials for high-temperature superconductors,
\cite{Kastner98}, 
or in noncentrosymmetric antiferromagnets.\cite{Zhelud97,PRB02}
However, the focus of current interest
in spin-flops has now shifted towards nanomagnetic systems.
Many of recently synthesized
nanostructured materials have magnetic
constituents with antiferromagnetic coupling. 
This vast and intensively
investigated class of \textit{antiferromagnetic}
nanostructures includes different types of
ferromagnetic/antiferromagnetic bilayers.\cite{Fitzsimmons00}
\textit{Synthetic} or \textit{artificial} antiferromagnets
are recently designed for
high-density storage technologies, spin-valves
and magnetic random access memory (MRAM) devices.
\cite{FullertonIEEE03,Maffitt06}
In addition to layered systems
nanoparticles of antiferromagnetic materials 
are currently investigated.
\cite{Hansen00}
A variety of \textit{size-dependent} electronic
and magnetic effects are intensively investigated
in many of these nanostructures.
\cite{bias,Mills94}
In particular, experiments
indicate that in ferro/antiferromagnetic bilayers 
the reorientation within the antiferromagnetic structures
strongly influences
the interface interactions and thus magnetic
properties of the
ferromagnetic subsystem.\cite{Nolting00,Nogues00}
A rich variety of specific field-induced
reorientational effects have
been found in synthetic antiferromagnets
such as toggle mode MRAM devices, \cite{Worledge04,APLNew} 
and finite antiferromagnetic superlattices.\cite{Mills94,PRB04}

In a broader context, 
new methods for the investigations of antiferromagnets
in the bulk or close to antiferromagnetic surfaces, 
\cite{Felcher96,Sievers,Stahl00} 
and of nanostructured antiferromagnets
\cite{Felcher96,Nogues00,Nolting00,Hansen00,Mills94}
provide a wealth of data 
on spin-reorientation effects.
These investigations require a refined 
theoretical description of these phenomena.
The theoretical understanding of the 
the spin-flop was developed during many 
years of investigations with contributions 
from many researchers and great progress 
towards an understanding of the magnetic
states near  the spin-flop transitions has 
been reached. 
Still, the present stage of the theory 
is mostly dedicated to separate phenomena 
and a coherent description of the magnetization
processes near spin-flops was not presented yet.
This problem is evidently one that should be addressed 
with phenomenological approaches of magnetism, 
i.e., a ``micromagnetism'' for antiferromagnetics.\cite{Hubert98}
First solutions were obtained for fields applied 
along the easy axis \cite{Turov65,Anderson64}. 
The stability limits forming ``critical astroids''
in oblique magnetic fields, i.e. in fields deviating 
from the easy-axis, and the ``critical angle 
of the spin-flop'' were calculated 
in Refs.~$[$\onlinecite{Rohrer69}$]$
for a model with second-order anistropy.
The influence of the fourth-order
anisotropy on the metastable states has been investigated 
in Ref.~$[$\onlinecite{Mitsek69}$]$. 
In Ref.~$[$\onlinecite{FTT82}$]$ the problem 
has been reduced
to an effective model akin to the usual reorientation 
in a uniaxial ferromagnet, and 
a complete set of solutions was derived.

In this paper we give a comprehensive 
phenomenological theory of the magnetic 
states and their evolution in applied
fields for a two-sublattice
collinear antiferromagnet.
The model is explained in section~\ref{model}.
We calculate all possible magnetic configurations in
the system (section~\ref{diagrams}) 
and give a physically clear description 
of the main features of the magnetization processes.
In particular we calculate and analyse the tensor of 
the static magnetic susceptibility (sections~\ref{magnetization} and
\ref{domains2}).
This enable us to generalize
and systematize results on bulk spin-flop 
and formulate directions 
for the investigation of this transition 
in bulk antiferromagnets and 
in antiferromagnetic nanostructures.
In section~\ref{domains2} the occurrence of 
multidomain states by demagnetization
effects is analysed.
There are characteristic peculiarities of the field
and angular dependencies of the magnetic susceptibility, 
which can be employed in experiments on the spin-flop transition.
In section~\ref{experiment} we demonstrate 
this approach based on experimental data 
for orthorhombic (C$_2$H$_5$NH$_3$)$_2$CuCl$_4$,
which is a layered model antiferromagnet. 
\cite{ZhurNEXT}


\section{Model and equations}\label{model}

Within  the phenomenological theory
of magnetism the magnetic (free) energy for
a bulk \textit{collinear} two-sublattices  
antiferromagnet can be written in the following form 
\cite{Turov65,UFN88}
\begin{eqnarray}
 W& =  & \int w({\mathbf m}_1, {\mathbf m}_2)d\,V  
\label{energy0}\\
&  = & \int\Bigg\{J{\mathbf m}_1\cdot{\mathbf m}_2 + 
 e_a({\mathbf m}_1, {\mathbf m}_2) \nonumber \\
& & -{\mathbf H}^{(\textrm{e})}\cdot({\mathbf m}_1 + {\mathbf m}_2)M_0^{-1} \nonumber\\
& & -\frac{1}{2}{\mathbf H}^{(\textrm{m})}\cdot({\mathbf m}_1 
+ {\mathbf m}_2)M_0^{-1} \Bigg\}M_0^2\,d\,V. \nonumber
\end{eqnarray}
We assume here that 
the vectors of the sublattice magnetizations $\mathbf{M}_j$
do not change their modulus  and their
orientations are
described by unity vectors  
$\mathbf{m}_j =\mathbf{M}_j/M_0$, $M_0 =|\mathbf{M}_j|$.
Hence, we develop our theory for 
constant low temperatures 
in the antiferromagnetically ordered state.
The energy (\ref{energy0}) consists of the  exchange
interaction with exchange constant$J$, 
the magnetocrystalline anisotropy energy $e_a$, 
and Zeeman energy contributions
due to the external magnetic field ${\mathbf H}^{(\textrm{e})}$ 
and the demagnetization field ${\mathbf H}^{(\textrm{m})}$, 
the latter giving the dipolar stray field energy. 
In this paper we are interested in antiferromagnets 
with a preferable direction of a magnetic ordering, 
i.e. \textit{easy-axis} systems. 
In uniaxial antiferromagnets the
\textit{easy-axis} coincides with 
the principal axis of symmetry, and
in orthorhombic crystals with 
one of the orthorhombic axis.
In this paper
we chose coordinates such 
that the easy axis is along the $z$-axis.
The energy density $e_a=e_a^{(u)}+e_a^{(b)}$ 
includes uniaxial $e_a^{(u)}$ and
in-plane $e_a^{(b)}$ parts. 
The uniaxial anisotropy $e_a^{(u)}$ consists of
invariants of type $m_{1z}^{2(n-k)}m_{2z}^{2k}$ 
where $n$ is an integer, and $k=1,2..n$.\cite{Turov65}
These energy contributions strongly decrease
with increasing $n$. 
In the vicinity of the spin-flop field 
the second-order uniaxial anisotropy 
is ``cancelled'' by the applied magnetic field. 
Hence, higher-order terms must be  
be taken into account.\cite{FTT82, UFN88}
Therefore, both second-order 
and fourth-order terms must be included 
in the theory for spin-flops.
One can write the uniaxial anisotropy as \cite{UFN88}
\begin{eqnarray}
e_a^{(u)}(m_{1z}, m_{2z})  & =& 
 -\frac{K}{2}(m_{1z}^2 + m_{2z}^2) - K' m_{1z}m_{2z}
\nonumber\\
& & -\frac{{K}_{20}}{4}(m_{1a}^4 + m_{2a}^4) \nonumber\\
& &- \frac{{K}_{21}}{4}m_{1a}m_{2z}(m_{1a}^2 + m_{2a}^2) \nonumber\\
& &- \frac{{K}_{22}}{2}m_{1a}^2 m_{2a}^2\,.
\label{uniaxial}
\end{eqnarray}
Usually, the second-order
terms with constants $K$, $K'$ 
play the dominant role for 
the orientation of the magnetic vectors. 
But, the fourth-order terms 
with $n=2$ and constants $K_{2k}$ 
become vital near the spin-flop,
as will be shown later.

The anisotropy energy in the basal plane
$e_a^{(b)}$ includes invariants
composed of in-plane components of the vectors 
$\mathbf{m}_i$ and, depending on the crystal symmetry,
includes quadratic terms for orthorhombic crystals
or terms of higher degree for uniaxial crystals. 
These energy contributions lead to additional 
orientational effects. They determine 
one preferable direction for orthorhombic symmetry
or more directions for uniaxial antiferromagnets 
for the spin orientation in the basal plane perpendicular
to the magnetic easy axis.

The model Eq.~(\ref{energy0}) with anisotropy
(\ref{uniaxial}) describes a vast group 
of antiferromagnetic crystals
with collinear order,
including such well-studied
compounds as MnF$_2$,  FeF$_2$, Cr$_2$O$_3$,
GdAlO$_3$ and others.\cite{UFN88}
In this class of antiferromagnets 
effects of magnetic couplings are absent 
that violate a \textit{collinear} 
and spatially homogeneous order in the ground state, 
such as competing exchange 
or Dzyaloshinkii-Moriya interactions.\cite{PRB02}

The equations minimizing 
the energy functional (\ref{energy0})
together with the Maxwell equations 
for the magnetostatic problem
determine the distributions of the magnetization fields
${\mathbf m}_i (\mathbf{r})$ 
and the stray field $\mathbf{H}^{(\textrm{m})}(\mathbf{r})$
in the sample.\cite{Hubert98}
These integro-differential equations 
are too complex and impractical, even for brute-force 
numerical calculations.
However, the problem can be reduced to a number
of simplified auxiliary problems.\cite{UFN88,Hubert98}
Following these standard methods we start
from the analysis of spatially homogeneous
states in \textit{fixed} internal magnetic fields
$\mathbf{H}= \mathbf{H}^{(\textrm{e})}+\mathbf{H}^{(\textrm{m})}$.

In the antiferromagnetic 
crystals displaying a spin-flop,
the exchange coupling is 
much stronger than the anisotropy energy, $J \gg w_a$.  
In this case it is convenient to use
the \textit{net magnetization} vector,
${\mathbf m}=({\mathbf m}_1+{\mathbf m}_2$)/2,
and the {\textit {staggered magnetization}} vector
(or vector of antiferromagnetic order),
${\mathbf l}=({\mathbf m}_1-{\mathbf m}_2$)/2,
as internal variables of the system.
Because $|{\mathbf m}_{i}|$ = 1
these vectors satisfy the constraints 
${\mathbf m}\cdot{\mathbf l}$ = 0, 
${\mathbf m}^2+{\mathbf l}^2 = 1$.\cite{Turov65}
Independent minimization of 
(\ref{energy0}) with respect to ${\mathbf m}$ 
yields (see Refs.~$[$\onlinecite{FNT86,PRB02}$]$ 
for details)
\begin{eqnarray}
\mathbf{m} = [\mathbf{H}
-\mathbf{n}(\mathbf{H}\cdot \mathbf{n})]/H_{\textrm{e}},
\qquad 
H_{\textrm{e}} = (2 J M_0)\,,
\label{m}
\end{eqnarray}
where $\mathbf{n}=\mathbf{l}/|\mathbf{l}|$ is the unity
vector along the staggered magnetization
and $H_{\textrm{e}}$ is the so-called exchange field. 
As follows from (\ref{m}), in this field $\mathbf{m}$ =1,
i.e. the magnetization vectors align along the field 
direction, which is a \textit{spin-flip} transition.
After substitution of (\ref{m}) and
omitting gradients of ${\mathbf m}$ the energy
density $w$ in Eq.~(\ref{energy0}) 
can be written as a function 
of the vector $\mathbf{n}$ alone
\begin{eqnarray}
w = -\frac{1}{2 J }[\mathbf{H}^2 -
(\mathbf{H}\cdot \mathbf{n})^2] 
+ e_a(\mathbf{n})M_0^{2}\,.
\label{energyw}
\end{eqnarray}

For the collinear antiferromagnets,
further simplifications are possible 
by restricting the spatial orientation 
of the  magnetization vectors.
In most practically important cases, 
the equilibrium configurations $\mathbf{m}_i$ and, 
correspondingly, the vectors 
$\mathbf{m}$ and $\mathbf{l}$ 
remain in or close to 
the plane spanned by the easy axis
and the magnetic field.\cite{FTT82} 
In this paper, we chose coordinates 
with $x0z$ as this plane.
The restriction on the magnetic configurations
always holds true
when the vector $\mathbf{H}$ remains in the plane spanned
by the ``easy'' and the ``intermediate'' axes 
of an orthorhombic antiferromagnet.
In the uniaxial antiferromagnets, 
in-plane components of the magnetic field usually 
suppress orientational effects of the weak 
in-plane anisotropy $w_a^{(b)}$. 
Then, the vectors $\mathbf{m}_i$ of 
equilibrium states remain in the $x0z$ plane.
This means that only the uniaxial anisotropy (\ref{uniaxial}) 
may essentially influence the magnetic states.
For the magnetic energy in terms 
of $\mathbf{m}$ and $\mathbf{l}$, 
a systematic analysis (see Ref.~$[$\onlinecite{FTT82}$]$)
shows that only the following terms 
from the uniaxial anisotropy (\ref{uniaxial})
must be retained,
\begin{eqnarray}
e_a^{(u)}=-(K+K')m_z^2-B_1l_z^2 - B_2l_z^4,\,
\label{uniaxial2}
\end{eqnarray}
where $B_1=K-K'$, 
$B_2=(K_{20}-K_{21}+K_{22})/2$.
Then, within the $x0z$  plane the magnetic states
are described by only one internal variable, the angle
$\theta$ between the easy axis and the staggered magnetization.
The expansion of the energy in terms of $\theta$
with respect to the small parameter $|e_a^{(u)}|/J$
yields the leading contribution
\begin{eqnarray}
w_0^{(1)}(\theta) = & & \label{energy01} \\ 
(4 J)^{-1} &&\!\!\!\!\left[
  \left(H_z^2- H_x^2- H_0^2 \right)\cos 2 \theta 
 + 2H_xH_z\sin 2 \theta \right]\,, \nonumber
\end{eqnarray}
\begin{eqnarray}
 H_0& =& \sqrt{2J B_1}M_0,\, 
\label{H0}
\end{eqnarray}
and the contributions to next order
\begin{eqnarray}
w_0^{(2)}(\theta) & = &  -\frac{H_0^2}{4J}
\left( \frac{B_2}{2B_1} + \frac{K}{2J} \right) \cos^2 2 \theta 
\nonumber\\
& & - \frac{H_0^2}{4J} \left( \frac{B_2}{2B_1} 
+ \frac{K-K'}{2J} \right)\cos 2 \theta \,.
\label{energy02}
\end{eqnarray}
In the transformation of the energy density (\ref{energyw}) 
into the simplified energy density $w_0(\theta) = 
w_0^{(1)}(\theta)+w_0^{(2)}(\theta)$ with the contributions
from Eqs.~(\ref{energy01}) and (\ref{energy02}),
we have omitted higher-order terms 
and a constant combination of materials constants.
The coefficients 
in $w_0^{(1)}$ in Eq.~(\ref{energy01})
are proportional to $H^2$. 
They are generally
much larger than those in $w_0^{(2)}$.
In wide regions of the magnetic phase-diagram 
$w_0^{(1)} \gg w_0^{(2)} $ and
the energy contribution $w_0^{(2)}$ can be neglected.
However, as the magnetic field approaches 
the critical point ($H_x,H_z$)=($0, H_0$) the leading
energy term in (\ref{energy01}) vanishes
and the term $w_0^{(2)}$ from Eq.~(\ref{energy02})
must be taken into account.
In next sections we analyze the magnetic states
for model (\ref{energy01}), (\ref{energy02}).

\begin{figure}
\includegraphics[width=8.5cm]{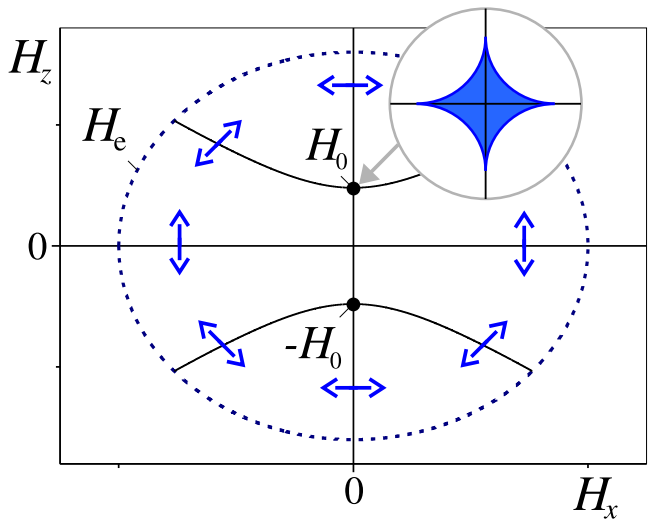}
\caption{
\label{NeelEquation}
($H_x, H_z)$ phase diagram for easy-axis antiferromagnets. 
The \textit{dashed} line $H_{\textrm{e}}$ gives the continuous transition 
into the saturated states. 
The shaded area is the region of the
metastable states in the vicinity of the critical field
($H_x =0, H_z =H_0$).
Details of the phases diagram in
these regions are shown in the next figure.
}
\end{figure}

\section{Magnetic phase diagrams}\label{diagrams}

\subsection{Equilibrium magnetic configurations}\label{phases2}

\begin{figure*}
\includegraphics[width=7.0cm]{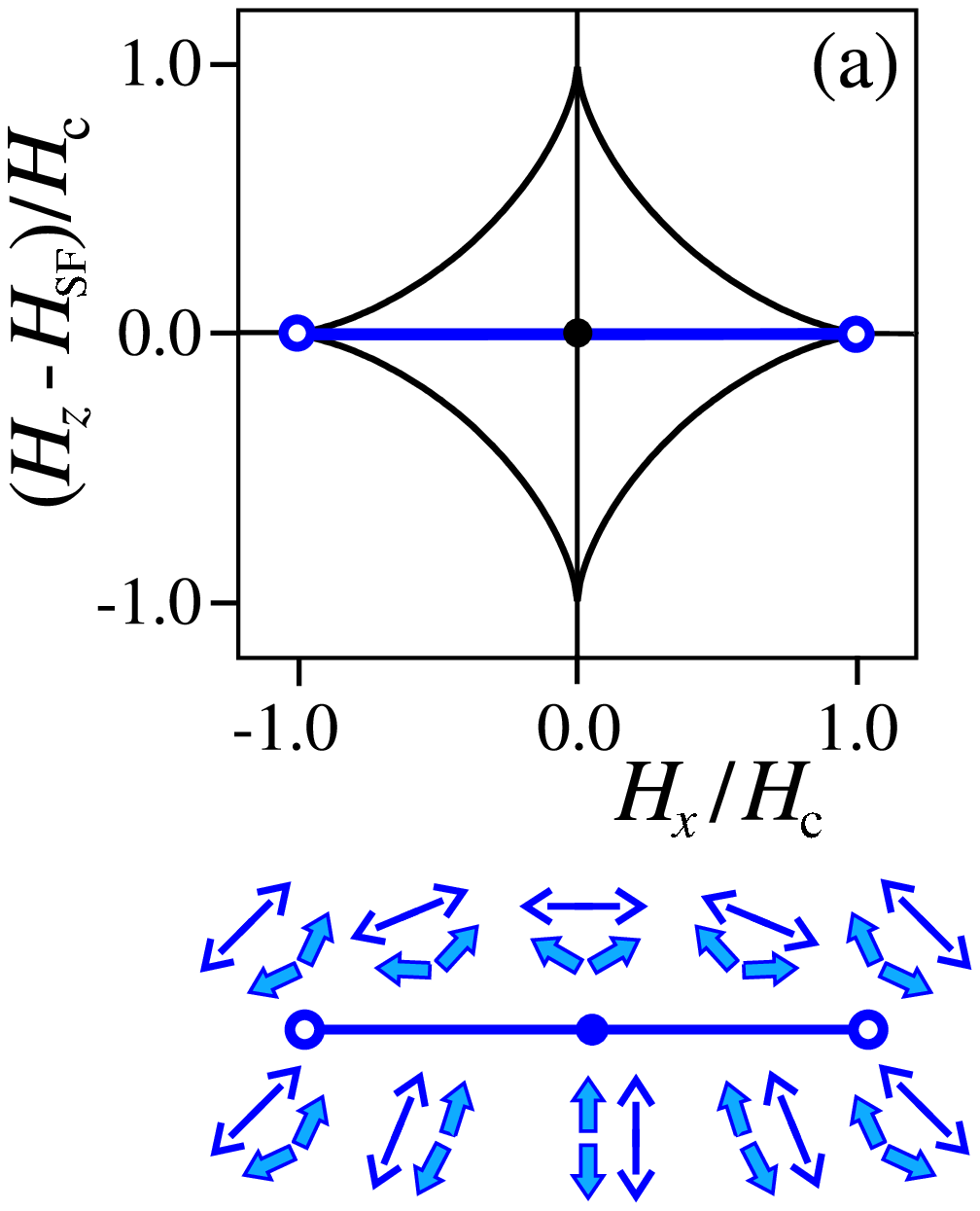}
\hspace{0.5cm}
\includegraphics[width= 9.8cm]{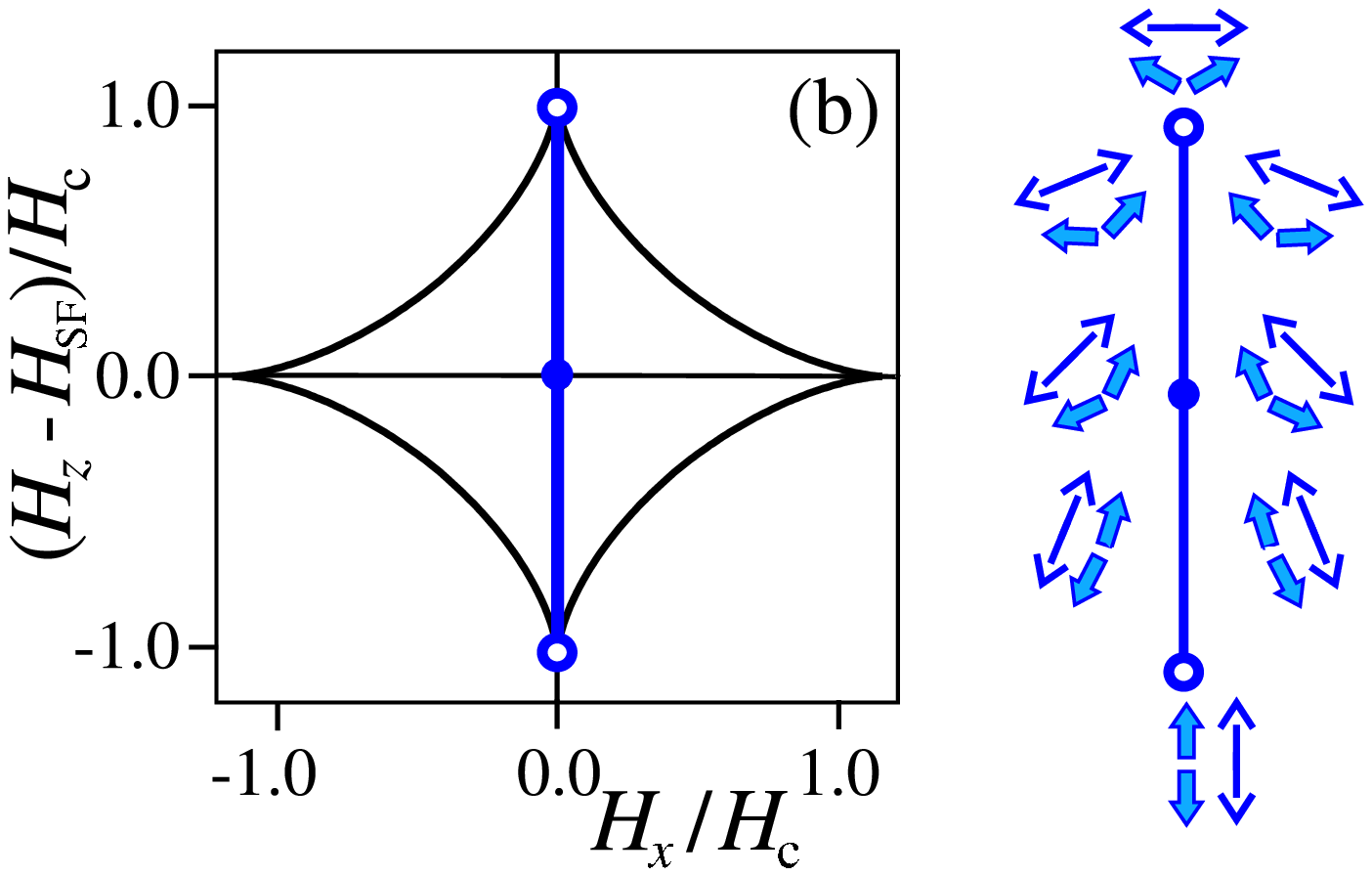}
\caption{
\label{HxHz}
Details of the phase diagram Fig.~\ref{NeelEquation}
in the vicinity of the spin-flop field depend
on the sign of $\varkappa$ 
as defined in Eq.~(\ref{energy3}):
(a) $\varkappa>0$, (b) $\varkappa<0$.
Thin lines confine the astroid regions 
with metastable states.
Thick line segments give
the first-order transition lines.
The arrows show spin-configurations
in the competing phases along the transition.
Hollow points are the end points of the
first-order transitions.
}
\end{figure*}
Outside the spin-flop region the equilibrium states
are described by the energy $w_0^{(1)}(\theta)$ (\ref{energy01}). 
%
%
In easy-axis antiferromagnets 
the staggered vector $\mathbf{l}$ interacts with
the applied magnetic field via the coupling to
the magnetization vector $\mathbf{m} \perp \mathbf{l}$
according to Eq.~(\ref{m}).
Correspondingly the interaction with the applied field
favours the state $\mathbf{l} \perp \mathbf{H}$.
Note that in isotropic systems one has $H_0 = 0$, and
the energy density  $w_0^{(1)}(\theta) = H^2\cos(2\theta-2\psi)$
with $\psi$ the angle between $\mathbf{H}$ and $z$-axis
has minima for $\theta=\psi \pm \pi/2$.
In easy-axis antiferromagnets 
the uniaxial anisotropy orientates
$\mathbf{l}$ along the easy axis. 
The competition between these
two magnetic couplings 
determines the equilibrium states.
These states are sketched in the ($H_x, H_z$) phase diagram,
Fig.~\ref{NeelEquation}.
Minimization of (\ref{energy01})
leads to  the well-known \textit{N\'eel formula} \cite{Neel36} 
for the magnetic configurations,
\begin{eqnarray}
\tan 2\theta = \frac{2H_{z}H_{x}}{H_{z}^2-H_{x}^2- H_0^2 }\,.
\label{NeelEq}
\end{eqnarray}
Because the energy is invariant under the transformation
$\theta \to \theta + \pi$,  Eq.~(\ref{NeelEq}) 
describes solutions with antiparallel directions 
of the staggered vector.
The equilibrium states $\theta_{1,2}$ correspond to 
the minima of the leading energy Eq.~(\ref{energy01}),
which are given by
$w_0^{(1)}(\theta_1) = w_0^{(1)}(\theta_2)
\equiv  \min [ w_0^{(1)}(\theta)] $. 
These wells in the potential 
energy  Eq.~(\ref{energy01}) 
are separated by a barrier. 
The \textit{height} of this potential barrier is
$\Delta w = \max [ w (\theta)] 
- \min [ w (\theta)] $.
For the potential (\ref{energy01}) we obtain
\begin{eqnarray}
\Delta w_0^{(1)} = 
\frac{1}{2J}\sqrt{ \left(H_z^2- H_x^2 - H_0^2 \right)^2
+ 4H_x^2 H_z^2 }\,.
\label{Neelbarrier}
\end{eqnarray}
In a magnetic field directed along the easy axis $H_x = 0$
the \textit{collinear} antiferromagnetic phase
for $H < H_0$
and the \textit{spin-flop} phase in the field range $H_0 < H < H_{\textrm{e}}$
correspond to the two branches of solutions for Eq.~(\ref{NeelEq})
with $\theta=0$  and $\theta=\pi/2$, respectively
(Fig.~\ref{phases}(a),(b)).
In magnetic fields that deviate from the easy axis,
\textit{angular} or \textit{canted} 
phases are realized (Fig.~\ref{phases}(c)).
The equilibrium states described 
by Eq.~(\ref{NeelEq}) result from the competition 
between the uniaxial anisotropy,
which favours the orientation of the
staggered magnetization $\mathbf{l}$ along the
easy axis and the magnetic field, 
which orientates $\mathbf{l}$ 
perpendicular to its direction.
At low fields the net magnetization is 
very small $|\mathbf{m}| \ll |\mathbf{l}|$,
so the energy contribution $B_1l_z^2 $
plays the dominating role for 
the orientation of the magnetic configuration.
The characteristic field $H_0$ given by a geometrical 
mean of the intra-sublattice exchange $J$ 
and the second-order anisotropy for the staggered vector $B_1$ 
measures the scale of the energy contributions 
favouring the easy-axis ground state. 
Thus, for small fields $|\mathbf{H}| < H_0$ 
the anisotropy prevails and stabilizes states
with the staggered vector  $\mathbf{l}$ 
nearly parallel to the easy-axis, $\theta \ll 1$,
with small magnetization, $m \ll 1$.

In an increasing field $\mathbf{l}$ rotates towards 
the direction perpendicular to $\mathbf{H}$, 
and the net magnetization $m$ gradually increases.
In the $(H_{x}, H_{z})$ phase diagram (Fig.~\ref{NeelEquation}) 
the hyperbola, $H_{z}^2- H_{x}^2= H_0^2$, separates
the regions with the angles $|\theta|$ 
larger and smaller $\pi/4$.
At $H = H_{\textrm{e}}$ the antiferromagnet transforms into the
saturated state with $m = 1$ along the applied field,
$\mathbf{m} || \mathbf{H}$.

The N{\'{e}}el equation Eq.~(\ref{NeelEq}) has a singularity 
at $(H_{x},H_{z})=(0,H_0)$. 
At this point $(0,H_0)$ in the phase-diagram
the main  competing forces completely compensate each other:
the leading energy contribution (\ref{energy01})
equals zero, and the next order energy contribution 
$w_0^{(2)}$ in Eq.~(\ref{energy02}) plays a decisive role.
In this region, the full energy of the model
$w^{(0)} (\theta))$
can be written as
\begin{eqnarray}
w_0 (\theta) & = & - \frac{\varkappa}{4J} H_0^2 \cos^2 2 \theta 
\label{energy3}\\
 & +& \frac{1}{4J}\left[
\left(H_z^2- H_x^2 - H_{\textrm{\textrm{SF}}}^2 \right)\cos 2 \theta \right.  
 \left.+2H_x H_z \sin 2 \theta \right]
 \nonumber
\end{eqnarray}
with
\begin{eqnarray}
\varkappa & = &  \frac{B_2}{2B_1} +\frac{K}{2J},\\
H_{\textrm{\textrm{SF}}} & = &  H_0 +\Delta H_0, \\
\Delta H_0 & = &  \sqrt{2JB_1}M_0\left(\frac{K + K'}{4J} 
+\frac{B_2}{2B_1} \right)\,.
\label{SFfield}	
\end{eqnarray}
Eq.~(\ref{energy3}) represents 
a consistent expression for the phenomenological energy 
near the SF field ($0, H_0$). 
It includes those higher-order interaction terms 
that are mandatory owing to the compensation of 
leading energy contributions (see Eq.~(\ref{energy01})).
These additional terms consist of inter- and 
intra-sublattice uniaxial second-order anisotropy
terms with parameters $K$, $K'$ and 
the fourth-order anistropy $B_2$ of the staggered vector.
Comparing energy densities $w_0^{(1)}$ from Eq.~(\ref{energy01}) with 
Eq.~(\ref{energy3})
we find that these magnetic coupling terms
(i) shift the value of the SF 
field $ H_0 \rightarrow H_{\textrm{\textrm{SF}}}= H_0 +\Delta H_0$,
where $\Delta H_0 \ll  H_0$, 
and 
(ii) create an additional interaction term  
proportional to $\cos^2 2 \theta$.
This energy contribution stabilizes
the magnetic states at the SF field ($0, H_{\textrm{\textrm{SF}}}$) because
it provides a potential barrier 
$\Delta w_0 (H_{\textrm{\textrm{SF}}}) = |\varkappa| H_0^2/(2J)$.

Due to the relations  $|K|,|K'|,|K_{2k}\|\ll J$, $|B_2|\ll |B_1|$
the region, where $w_0^{(1)}$ and $w_0^{(2)}$ have the same
order, 
is restricted to the close vicinity of
the point ($H_x=0, H_z= H_0$):
$|H_z-H_0| \ll H_0, |H_x| \ll H_0 $.
In this region the energy density
$w_0(\theta)= w_0^{(1)}(\theta)+w_0^{(2)}(\theta)$
can be reduced to the potential expression
\begin{eqnarray}
\Phi(\theta)  =  \frac{2 J w_0}{H_0}  &=& 
- \frac{\textrm{sign}({\varkappa})}{2}H_{\textrm{c}} \cos^2 2 \theta 
\label{Phi4} \\
& +& 
\left(H_z-H_{\textrm{\textrm{SF}}} \right) \cos 2 \theta 
+ H_x \sin 2 \theta\,\nonumber. 
\end{eqnarray}
Here, the characteristic field 
\begin{eqnarray}
H_{\textrm{c}} = |\varkappa| H_0 = 
\left|\frac{B_2}{2B_1} +\frac{K}{2J} \right|\sqrt{2JB_1}M_0 \ll H_0
\label{Hc}
\end{eqnarray}
sets the scale of the field region 
around the critical field (0,$H_0$), 
where the interactions described 
by the energy density (\ref{Phi4}) 
produce noticeable effects.
The energy density $\Phi (\theta)$ (\ref{Phi4}) 
functionally 
coincides
with that of a uniaxial ferromagnet
\begin{eqnarray}
\Phi_f(\phi) = 
-\frac{1}{2}\beta M_0  \cos^2 \phi
-H_z \cos  \phi
- H_x \sin  \phi\,,
\label{PhiF}
\end{eqnarray}
where $\beta$ is the anisotropy constant,
$\phi$ is the angle between the magnetization
$\mathbf{M}$ and the $0z$ axis.
\cite{Hubert98}
In the ferromagnetic model (\ref{PhiF})
the anisotropy field $H_a =|\beta|M_0$
corresponds to the characteristic field
$H_{\textrm{c}}$ in (\ref{Phi4}),
the components of the magnetic field
($H_x, H_z$) correspond to
($-H_x, -(H_z-H_{\textrm{SF}}$),
and the angle
$\phi$ to the angle $2 \theta$.
Thus, the behaviour of the uniaxial antiferromagnet is
reduced to the well-known model and the corresponding 
mathematical results for the magnetic states of
uniaxial ferromagnets.\cite{Stoner48,Hubert98}

For the model (\ref{Phi4}) the equation for the equilibrium state 
$d \Phi / d \theta =0$ is
\begin{eqnarray}
\textrm{sign}(\varkappa) H_{\textrm{c}} \sin 2\theta\cos 2\theta  &&
\label{Eq}\\
 - (H_z -H_{\textrm{SF}})\sin 2\theta & + & H_x \cos 2\theta  =  0 \nonumber
\end{eqnarray}
and the existence region for metastable
states is bounded by the astroid 
\begin{eqnarray}
H_{x}^{2/3} + \left(H_{z} -H_{\textrm{SF}} \right)^{2/3} = H_{\textrm{c}}^{2/3}\,.
\label{astroid}
\end{eqnarray}
The polar angle  for the cusps on the sides of the astroid 
\begin{eqnarray}
\psi_{\textrm{c}} = \arctan (|H_{\textrm{c}}|/H_0)= 
\arctan \left(\varkappa \right) \approx \varkappa,
\label{PsiKrit}
\end{eqnarray}
is the so-called \textit{critical angle of the spin-flop}. 
This is the maximal angle for the existence of 
the metastable states in obliquely applied magnetic fields.
The notion of this ``critical angle''
was introduced in Ref.~$[$\onlinecite{Rohrer69}$]$, 
where the astroids of type (\ref{astroid}) have 
been calculated for the model with second-order anisotropy.
The character of the magnetic states 
within the astroid (\ref{astroid})
depends on the sign of the 
parameter  $\varkappa$ (\ref{energy3}):

\noindent
(i) For $\varkappa >0$
in magnetic field $H_{z} = H_{\textrm{SF}}$, $H_{x}=0$, the first-order
transition occurs between the antiferromagnetic and spin-flop phase.
This is the proper 
jump-like \textit{spin-flop transition}.
Note that the characteristic field
$H_0$ in the N\'eel equation (\ref{NeelEq}) 
differs from the spin-flop field
$H_{\textrm{SF}}$, Eq.~(\ref{SFfield}).
At finite transversal components of the magnetic field $H_{x}$ 
the first-order transition happens between distorted AF and SF,
i.e., canted phases 
at the line ($H_{z}= H_0, |H_{x}| \le \varkappa$) 
(Fig.~\ref{HxHz}(a)).
The solutions for these competing phases 
are (cf. Ref.$[$~\onlinecite{FTT82}$]$)
\begin{eqnarray}
\theta_1=-\frac{1}{2}\arcsin(H_{x}/H_{\textrm{c}}), \quad
\theta_2=-\pi/2 - \theta_1\,.
\label{solution1}
\end{eqnarray}
For increasing $H_{x}<H_{\textrm{c}}$ the difference 
between the competing canted phases
decreases. This difference disappears at 
endpoints of the first-order transition lines. 
These endpoints are located at the cusps of the astroid
(\ref{astroid}), $(H_{\textrm{c}}, H_{\textrm{SF}})$, $(-H_{\textrm{c}}, H_{\textrm{SF}})$.
The configuration of the magnetic states in these points 
is for both phases $\theta_{1,2}=\mp\pi/4$ (Fig.~\ref{HxHz}(a)). 

\noindent
(ii) For $\varkappa < 0$,
the canted phases exist as stable states even 
in magnetic fields along the easy axis
within the astroid (\ref{astroid}).
Minimization of (\ref{energy3}) for
this case yields 
the deviation angle of the solutions 
from the easy-axis
\begin{eqnarray}
\theta_{b}= \pi-\arccos[(H_{z}-H_0)/|H_{\textrm{c}}|]\,.
\label{solution2}
\end{eqnarray}
These solutions describe a continuous rotation of 
the staggered vector $\mathbf{l}$ from the AF phase
at the low cusp, $H_z =H_0 -|H_{\textrm{c}}|$, to the SF phase 
at the high cusp, $H_z =H_0+ |H_{\textrm{c}}|$ (Fig.~\ref{HxHz}(b)). 
The plane(s) of this rotation are determined 
by in-plane magnetic anisotropy.
Depending on the crystal symmetry there are several such
planes (half-planes) spanned by the easy-axis and easy magnetization
directions in the basal plane. 
This degeneracy of the magnetic states is lifted
by a deviation of the applied field from the easy axis.
In such a field, the canted state 
with the largest projection of $\mathbf{m}$ 
onto the field direction 
corresponds to the stable states, 
while other states preserve metastability for small 
deviations of the field and 
become unstable for larger deviations. 
This means that the vertical axis of the astroid (\ref{astroid})
given by $H_0-|H_{\textrm{c}}| < H_z <H_0+|H_{\textrm{c}}|$ 
is a first-order transition
line between several canted phases.
In particular, in orthorhombic antiferromagnets there is only
one plane of rotation (spanned by the easy and intermediate axis) 
and the phase transition occurs between two canted
phases with opposite rotation sense. Such magnetic states
have been studied in Ref.~$[$\onlinecite{FTT82}$]$.

In next subsections we analyze field 
dependencies of the magnetization
and the magnetic susceptibility.

\subsection{Magnetization and magnetic susceptibility}\label{magnetization}

\begin{figure*}
\includegraphics[width=8.5cm]{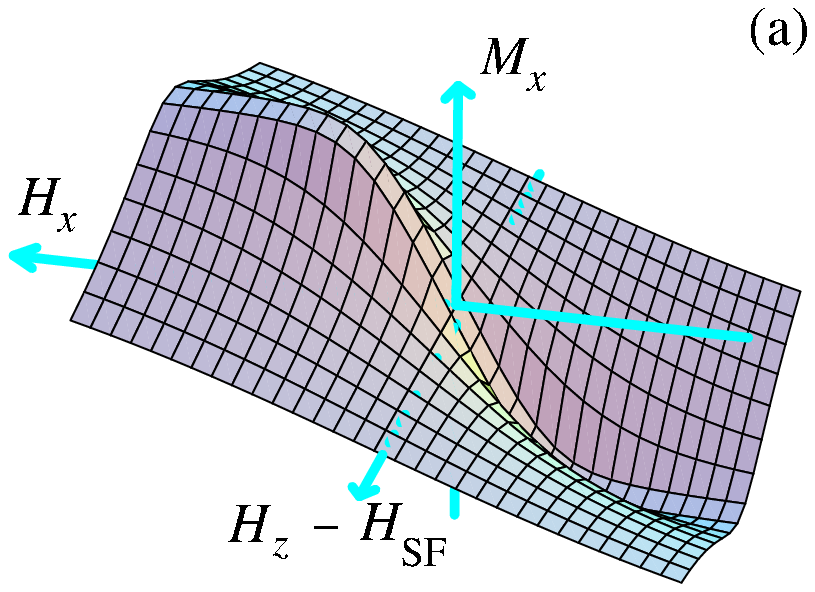}
\includegraphics[width=8.5cm]{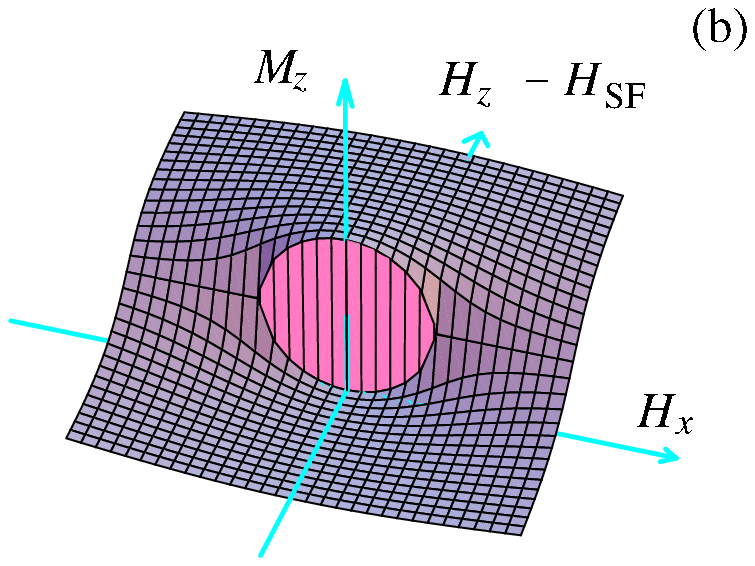}
\includegraphics[width=8.5cm]{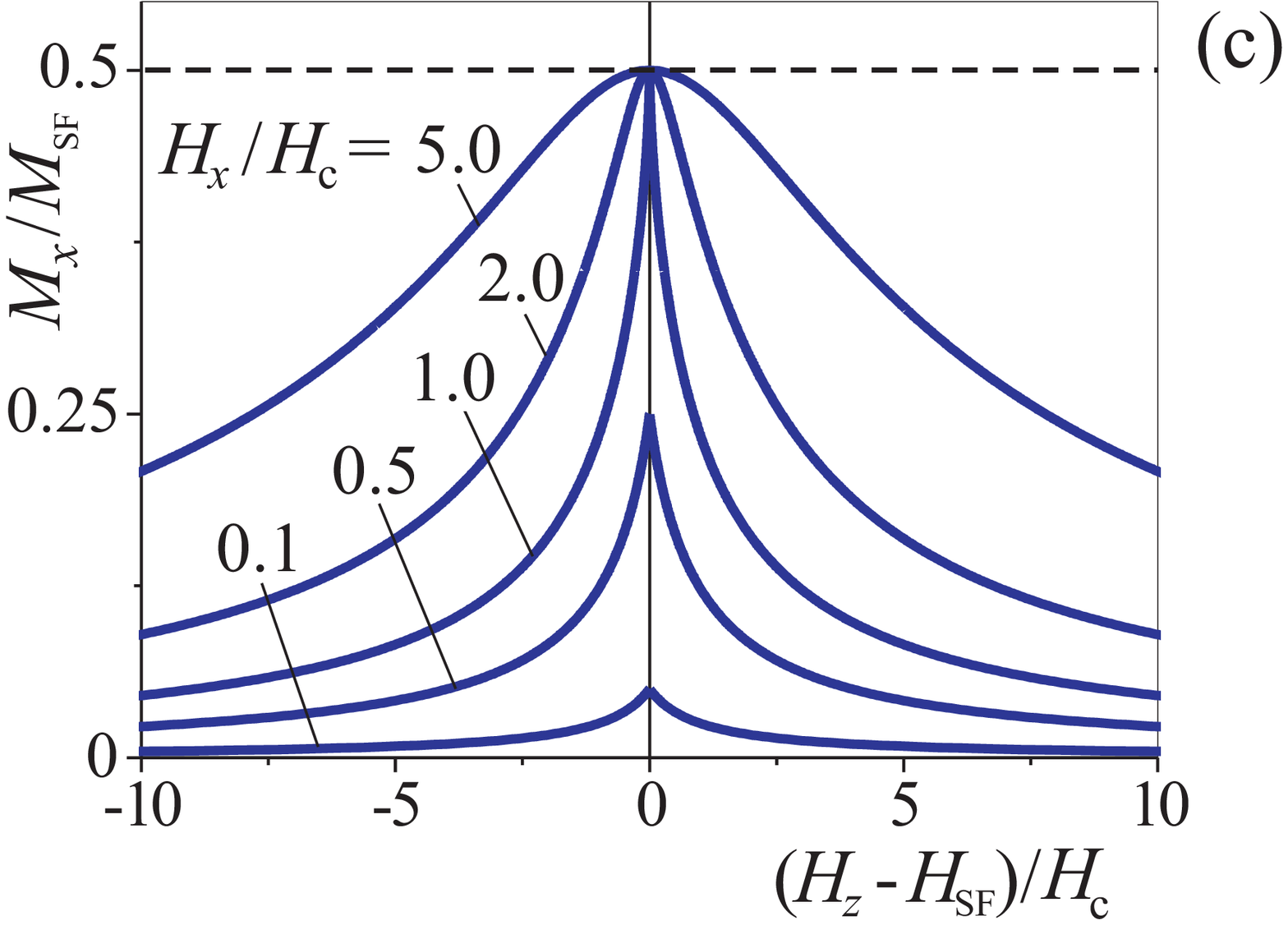}
\includegraphics[width=8.5cm]{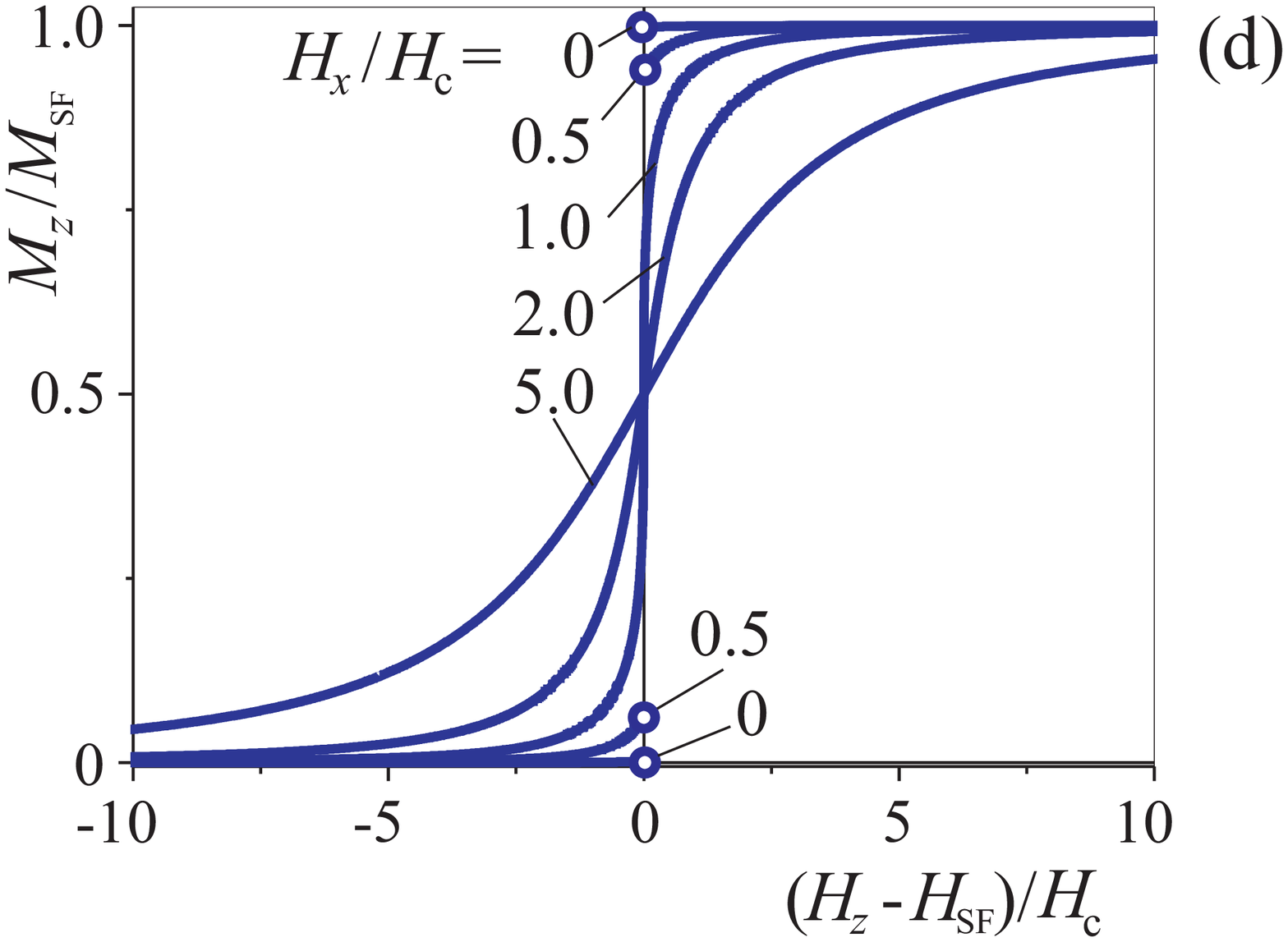}
\caption{
\label{magnetizationSF}
Field dependencies of the magnetization components
near the SF field for $\varkappa > 0$:
transverse $M_x (H_x, H_z)$(a),
longitudinal $M_z (H_x, H_z)$ (b)
and their projections onto the ($H_z,M_{ x (z)}$) 
planes (c), (d).
Only the magnetization curves corresponding to
the stable states are plotted.
}
\end{figure*}

For the field dependencies of the net magnetization 
$\mathbf{M} = M_0(\mathbf{m}_1 + \mathbf{m}_2) = 2M_0\mathbf{m}$,
one derives from (\ref{m})
\begin{eqnarray}
M_x = \frac{M_0}{2J}
\left[H_x(1+\cos 2\theta) - H_z \sin 2 \theta \right]
\nonumber\\
M_z = \frac{M_0}{2J}
\left[H_z(1-\cos 2\theta) - H_x \sin 2 \theta \right]\,,
\label{Mi}
\end{eqnarray}
where $\theta$ is the solutions of the 
equations (\ref{NeelEq}) and (\ref{Eq})
minimizing the energy. 
For the calculation of the magnetization vector 
and susceptibility tensor in the vicinity of
the SF field, it is convenient to rewrite 
the components of the total magnetization (\ref{Mi}) 
in the following form 
\begin{eqnarray}
M_x  & = &  M_{\textrm{SF}} 
\left[-\sin 2 \theta  + \mu_1 (\theta) \right] , \nonumber\\
M_z  & = & M_{\textrm{SF}}\left[ \left( 1 
- \cos 2 \theta \right)+ \mu_2 (\theta)\right]\,,
\label{m3}
\end{eqnarray}
where
\begin{eqnarray}
\mu_1(\theta) & = & 
\frac{|\varkappa|}{H_{\textrm{c}}} 
\left[H_z \sin 2 \theta + H_x(1 + \cos 2 \theta)\right],\nonumber\\
\mu_2(\theta)  &=& 
\frac{|\varkappa|}{H_{\textrm{c}}} 
\left[ H_z(1 - \cos 2 \theta) - H_x \sin 2 \theta \right]\,.
\label{mu}
\end{eqnarray}
The magnetization 
\begin{eqnarray}
M_{\textrm{SF}} = M_0\sqrt{\frac{B_1}{2J}} = M_0 \left(\frac{H_0}{H_{\textrm{e}}}\right)
 \label{msf}
\end{eqnarray}
characterizes the typical values of the net
magnetization in the spin-flop region.
By virtue of the relation $H_{\textrm{e}} \gg H_0$
the magnetization $M_{\textrm{SF}}$ amounts 
only to a small fraction
of the saturation value $M_0$.

In the SF region, given by  $H_x, H_z \lesssim H_{\textrm{c}}$,
see Eq.~(\ref{Phi4}),
the functions  $\mu_1$ and $\mu_2$ are very small,
$\mu_i \ll 1, i=1,2$, and can be omitted.
By substituting the solutions (\ref{solution1}) into (\ref{Mi}) 
we obtain  the magnetization on the transition line
for $\varkappa > 0$,
\begin{eqnarray}
M_{x}^{(1)} & = & M_{x}^{(2)}=\frac{M_{\textrm{SF}}}{2}
\left(\frac{H_x}{H_{\textrm{c}}} \right), \nonumber\\
M_{z}^{(1,2)} & = & \frac{M_{\textrm{SF}}}{2}\left[1 \mp \sqrt{1-
\left(\frac{H_x}{H_{\textrm{c}}} \right)^2}\right].  
\label{mP1}
\end{eqnarray}
The transverse components $M_x$ are equal in both phases.
The longitudinal components undergo 
a jump $ \Delta M_z =M_{\textrm{SF}} \sqrt{1-
(H_x/H_{\textrm{c}})^2}$ at the SF (Fig.~\ref{magnetizationSF}).
The parameter $M_{\textrm{SF}}$ Eq.~(\ref{msf}) is equal
to the maximum value of the magnetization jump
at the SF transition.

For $\varkappa < 0$ 
the magnetization in the competing phases 
have antiparallel components perpendicular 
to the easy axis, while the parallel 
components are equal
\begin{eqnarray}
M_x^{(1,2)} & =  & \mp \frac{M_{\textrm{SF}}}{2}
\sqrt{1-\left(\frac{H_z-H_{\textrm{SF}}}{H_{\textrm{c}}} \right)^2}, \nonumber\\
M_z^{(1)}& = & M_z^{(2)}=\frac{M_{\textrm{SF}}}{2}\left[1 
+ \left(\frac{H_z-H_{\textrm{SF}}}{H_{\textrm{c}}} \right) \right].
\label{mP2}
\end{eqnarray}
In this case the transverse components $M_x$ 
undergo a jump given by
$ \Delta M_x =M_{\textrm{SF}} \sqrt{1-
(H_z-H_{\textrm{SF}})^2/H_{\textrm{c}}^2}$  at the first-order transition.
\begin{figure}[tbh]
\includegraphics[width=8.5cm]{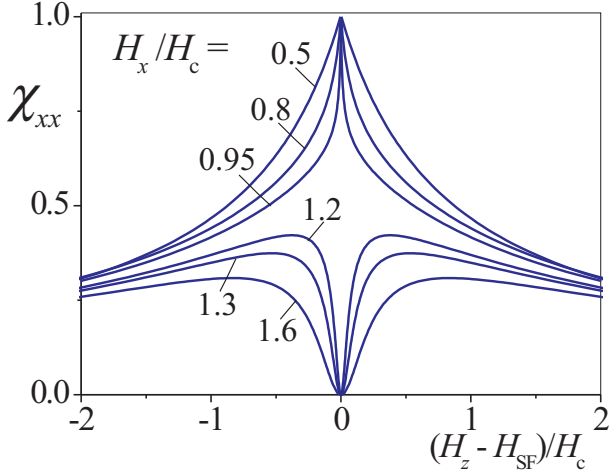}
\caption{
\label{Fchixx}
Field dependencies of $\chi_{xx}(H_z)$
for various values of $H_x$.
This and in two following figures display 
susceptibility components for antiferromagnets 
with $\varkappa > 0$. The susceptibility units are $1/(4J\varkappa)$.
Only the branches corresponding to 
the stable states are plotted.
}
\end{figure}
\begin{figure}[bth]
\includegraphics[width=8.5cm]{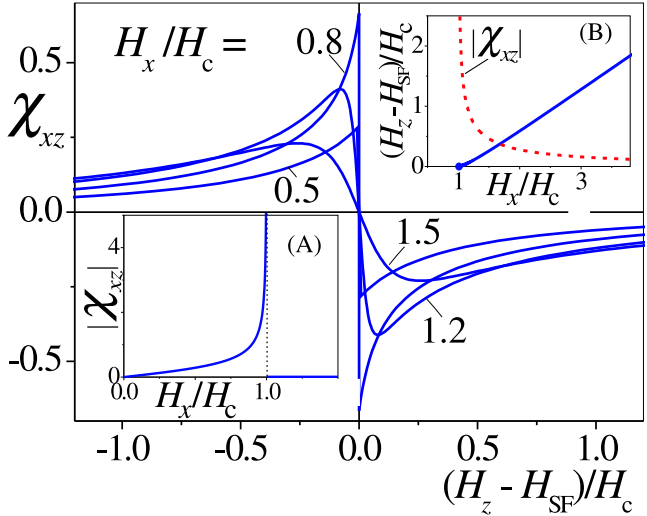}
\caption{
\label{Fchixz}
Field dependencies of $\chi_{xz}(H_z)$.
Inset (A) shows  $\chi_{xz}(H_x)$ at
the transition line $H_z = H_{\textrm{SF}}$.
In Inset (B), the location of the extremal values 
from Eq. (\ref{MaxXZ1}) are plotted by a solid line 
and their amplitudes from Eq. (\ref{MaxXZ2}) by
a dashed line.
}
\end{figure}
\begin{figure}[bht]
\includegraphics[width=8.5cm]{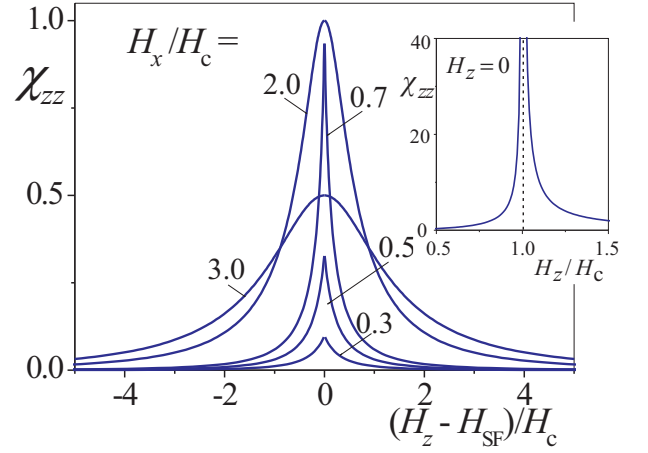}
\caption{
\label{Fchizz}
Field dependencies of $\chi_{zz}(H_z)$
for various values $H_x$.
The arrow-like shape of the functions
$\chi_{xx}(H_z)$ within the transition
region ($|H_x| < H_{\textrm{c}}$) changes into
a bell-like shape outside this region.
In both cases the functions $\chi_{zz}(H_z)$ 
have a maximum at $H_z = H_{\textrm{SF}}$.
These maximum values are plotted 
as function of $H_x$ in the inset.
}
\end{figure}

The landscapes of the magnetization ``surfaces''
$M_x(H_x,H_z)$, $M_x(H_x,H_z)$  
near the first-order transition and in the adjacent 
regions of the phase-space are shown in Fig.~\ref{magnetizationSF}.
The shape of these surfaces is reflected 
by peculiarites of the magnetic susceptibility.
The components of the tensor of
the \textit{internal} static magnetic susceptibility,
$\chi_{ij}^{(\textrm{i})}= \partial M_i / \partial H_j$,
are derived from the equations (cf.~$[$\onlinecite{JMMM05}$]$)
\begin{eqnarray}
 & & \chi_{xx}^{(\textrm{i})}=\frac{1}{4J}
[ 1 + \cos 2 \theta + 2H^2 \Omega(\theta) \cos^2 (2 \theta - \psi)],
\nonumber\\
& &  \chi_{xz}^{(\textrm{i})}= \frac{1}{4J}
[  \sin 2 \theta + H^2 \Omega(\theta) \sin (2 \theta - \psi)\cos (2 \theta - \psi) ]
\nonumber\\
& & \chi_{zz}^{(\textrm{i})}=\frac{1}{4J}
[ 1 - \cos 2 \theta + 2H^2 \Omega(\theta) \sin^2 (2 \theta - \psi)],
\label{chi}
\end{eqnarray}
where $\Omega^{-1}(\theta)=d^2 \Phi(\cos 2 \theta)/d (\cos 2 \theta)^2$,
$\Phi(\cos 2 \theta)$ is derived from Eq.~(\ref{Phi4}).
These relations, 
together with Eq.~(\ref{Eq}) 
yield field dependencies of the magnetization components 
and the susceptibility tensor in the spin-flop region.
Near the SF field the expansion of
(\ref{chi}) with respect to 
the small parameter $\varkappa  \ll 1$
allows a considerable simplification of the
expressions for $\chi_{ij}$,
\begin{eqnarray}
\chi_{xx}&  = & 
- \frac{1}{4J|\varkappa| } \; 
\frac{ \sin 2 \theta \cos^2 2 \theta}
{\left(\textrm{sign} (\varkappa) \sin^3 2\theta + H_x/H_{\textrm{c}} \right)} \nonumber\\
& = & 
- \frac{1}{4J|\varkappa| }  \frac{\cos^3 2 \theta}
{\left( \textrm{sign} (\varkappa) \cos^3 2\theta - H_z/H_{\textrm{c}} \right)}\,,
\label{chixx}
\end{eqnarray}
\begin{eqnarray}
\chi_{xz} & = & 
\frac{1}{4J|\varkappa|} \; \frac{
\sin^2 2 \theta \cos 2 \theta}
{\left(\textrm{sign} (\varkappa) \sin^3 2 \theta + H_x/H_{\textrm{c}} \right)} \nonumber\\
& = &  
- \frac{1}{4J|\varkappa|} \; \frac{\sin 2 \theta \cos^2 2 \theta}
{\left(\textrm{sign} (\varkappa) \cos^3 2 \theta - H_z/H_{\textrm{c}} \right)}\,,
\label{chixz}
\end{eqnarray}
\begin{eqnarray}
\chi_{zz} & = & 
- \frac{1}{4J|\varkappa|} \; \frac{\sin^3 2 \theta }
{\left(\textrm{sign} (\varkappa) \sin^3 2 \theta + H_x/H_{\textrm{c}} \right)} \nonumber\\
& = & 
\frac{1}{4J|\varkappa|} \; \frac{\sin^2 2 \theta \cos 2 \theta}
{\left( \textrm{sign} (\varkappa) \cos^3 2 \theta - H_z/H_{\textrm{c}} \right)} \,.
\label{chizz}
\end{eqnarray}
For antiferromagnets with $\varkappa > 0$ 
the expressions 
from Eqs.~(\ref{chixx}), (\ref{chixz}), and (\ref{chizz})
for $\hat{\chi}_{ij}$ in dependence on $H_x$ are convenient for
numerical calculations and 
those depending on $H_z$ in the case $\varkappa < 0$.
As the field approaches 
the stability limits, 
given by the lines $\Phi_{\theta \theta} \equiv \partial^2\Phi/\partial \theta^2 \rightarrow 0$,
the  functions $\chi_{ij} \propto 1/\Phi_{\theta \theta}$ diverge.
For applications to real antiferromagnets
it is important to distinguish the branches
of the functions $\chi_{ij}(H_x, H_z)$ 
corresponding to the stable states.
Typical field dependencies of the components
of $\hat{\chi}_{ij}$ for the stable
phases are plotted 
in Figs.~\ref{Fchixx}, \ref{Fchixz}, and \ref{Fchizz}.
The functions $\chi_{xx}(H_z)$
have \textit{qualitatively} different
field dependencies within 
and outside the region near the first-order transitions. 
In the former case $|H_x| < H_{\textrm{c}}$, the curves
$\chi_{xx}(H_z)$ have arrow-like shape with a 
maximum at the transition field $H_z = H_{\textrm{SF}}$.
In the latter case $|H_x| > H_{\textrm{c}}$, 
the curves $\chi_{xx}(H_z)$ have a minimum
in fields along a line prolongating the 
transition line and two symmetric maxima
\begin{eqnarray}
\chi_{xx}^{(\max)}= \frac{1}{4J \varkappa} 
\left( 1 - \frac{1}{3 \cos^2 \alpha} \right)
\label{MaxXX1}	
\end{eqnarray}
for fields lower and higher than the SF field $H_{\textrm{SF}}$.
The location of these maxima in ($H_x, H_z$)
for $H_x > H_{\textrm{c}}$  is determined parametrically by
a set of equations
\begin{eqnarray}
H_x^{(\max 1)} &  = & \frac{2 H_{\textrm{c}} \cos^3 \alpha }{3 \cos^2 \alpha -1}, \nonumber\\
H_z^{(\max 1)} &  = & \pm\frac{H_{\textrm{c}} \sin^3 \alpha }{3 \cos^2 \alpha -1}
\label{MaxXX2}	
\end{eqnarray}
with $\alpha = -\pi/2 - 2\theta$.
For $\chi_{xz}$ (Fig.~\ref{Fchixz}) stationary points
\begin{eqnarray}
|\chi_{xz}^{(\max)}|= \frac{1}{8J \varkappa} 
\frac{ 3 \cos^2 \alpha -2}{3|\sin \alpha| \cos \alpha}
\label{MaxXZ1}	
\end{eqnarray} 
are situated on  the lines
\begin{eqnarray}
H_x^{(\max 2)} &  = & \frac{ H_{\textrm{c}} \cos^3 \alpha }{3 \cos^2 \alpha -2}, \nonumber\\
H_z^{(\max 2)} &  = & \pm\frac{2H_{\textrm{c}} \sin^3 \alpha }{3 \cos^2 \alpha -2}\,.
\label{MaxXZ2}	
\end{eqnarray}
Similarly, functions 
$\partial \chi_{ij} / \partial H_k (\mathbf{H})$
also show qualitatively different dependencies
in different parts of the phase diagram
and include extremal points. 
The field values for all these anomalies 
are determined by the material constants of 
the antiferromagnets.
This connection between anomalous field-dependence
of the susceptibility tensor in the SF region and materials paramters 
can provide a basis for experimental approaches 
to investigate the magnetic 
properties and materials parameters in antiferromagnets.
Values of magnetic interactions, in particular
higher-order contributions to the magnetic anisotropy,
can be determined by measurements of the 
susceptibility tensor in the SF region 
and fitting of the $\chi_{ij}$ data 
to the theoretical expressions shown above.
This will be demonstrated in  section~\ref{experiment}.

Here, we compare characteristic values
of the magnetic susceptibility for
antiferromagnets in the major
part of the magnetic field phase diagram 
and in the SF region.
In  the ($H_x, H_z$) phase diagram 
(Fig.~\ref{NeelEquation}) saturation
is achieved at the exchange field $|H|=H_{\textrm{e}}$.
Thus, the average susceptibility is
$\left\langle \chi \right\rangle_{AFM}  = M_0/H_{\textrm{e}} = 1/(2J)$.
This is exactly the value of the susceptibility 
in the SF phase.
The metastable region in the phase diagram 
near the SF field has a width
$\Delta H = 2H_{\textrm{c}}=2H_0 \psi_{\textrm{c}}$, Eq.~(\ref{astroid}),
while the magnetization changes by $\Delta M = M_0 H_0/H_{\textrm{e}}$,
Eqs.~(\ref{mP1}), (\ref{mP2}). 
Thus, the average susceptibility in this region 
can be estimated to $\left\langle \chi \right\rangle_{\textrm{SF}}  
= M_0/(2|\varkappa|H_{\textrm{e}}) = 1/(4J\varkappa) \simeq 1/(4J\psi_{\textrm{c}})$.
This average equals the expressions 
from Eqs.~(\ref{chixx}), (\ref{chixz}), and (\ref{chizz}) up 
to some numerical factor.
The magnetic susceptibility 
near the SF field is strongly enhanced
compared to the average susceptibility in the 
major part of the phase diagram,
$\left\langle \chi \right\rangle_{\textrm{SF}}  
= \left\langle \chi \right\rangle_{AFM} (H_0/(2H_{\textrm{c}}))$.
This enhancement is given by the ratio $(H_0/(2H_{\textrm{c}})) = 1/(2\psi_{\textrm{c}})$
which usually amounts to a factor of several hundreds.
Note that the absolute change of the magnetization 
$\Delta M = M_0 H_0/H_{\textrm{e}} \ll M_0$ is tiny.
However, due to the extremely narrow 
width of the metastable region near the 
SF field, the magnetic susceptibility 
becomes very strong in this region.

As was mentioned above, 
similar functional expressions describe the field dependence of 
the magnetization $\mathbf{M}^{(f)}$ for the model of a uniaxial 
ferromagnet (\ref{PhiF})
and for the antiferromagnet in the SF regions 
following the basic approach Eq.~(\ref{m3}) with $\mu_i = 0$.
In fact, one can derive
\begin{eqnarray}
\mathbf{M}(\mathbf{H}/H_{\textrm{c}}) =\!\!\left(\frac{H_0}{2H_{\textrm{e}}} \right)\!
\mathbf{M}^{(f)} (\mathbf{H}/H_a) + 
M_0\!\!\left(\frac{H_0}{2H_{\textrm{e}}} \right)\!\mathbf{a}.
\label{ratioM}	
\end{eqnarray}
The last term in (\ref{ratioM}) signifies 
a shift in direction of the easy axis $\mathbf{a}$.
A corresponding relation 
between the components of the magnetic
susceptibility near the SF transition
and those of the ferromagnetic
susceptibility $\chi_{ij}^{(f)}$ 
is given by
\begin{eqnarray}
\chi_{ij}(\mathbf{H}/H_{\textrm{c}}) = 
\left(\frac{H_a}{2H_{\textrm{e}} |\varkappa|} \right)
\chi_{ij}^{(f)} (\mathbf{H}/H_a)\,.
\label{ratioChi}	
\end{eqnarray}
These relations demonstrate the 
physical similarity of 
the field dependencies for the magnetic properties 
in uniaxial ferromagnets 
and in easy-axis antiferromagnets near the SF field.
This equivalence is established by introducing reduced units 
and a shift for the magnetization in Eq.~(\ref{ratioM}),
and by reduced units and a scale-factor for the susceptibility 
in Eq.~(\ref{ratioChi}).
It is a consequence of the formal similarity 
in the phenomenological models Eqs.~(\ref{PhiF}) 
and (\ref{m3}) for both systems.
The relations  Eqs.~(\ref{ratioM}) and (\ref{ratioChi}) 
are useful for comparative studies 
on reorientation transitions in uniaxial ferromagnets 
and antiferromagnets near the spin-flop.

\subsection{Two-scale character of easy-axis antiferromagnets}\label{twoscale}

The magnetic states analyzed 
in the previous two sections display the pronounced 
\textit{two-scale} character of easy-axis antiferromagnets.
The magnetic phase diagrams in Fig.~\ref{HxHz}
comprise the main features
of the solutions in the spin-flop region.
On the large scale of the ($H_x, H_z$) phase diagram 
two materials parameters rule the behaviour:
the exchange or saturation field $H_{\textrm{e}}$ 
and the SF field $H_0$ 
defined in Eqs.~ (\ref{m}) and (\ref{H0}), respectively
(Fig.~\ref{NeelEquation}).
The former characterizes 
the strength of the antiferromagnetic 
exchange interaction,
the latter comprises 
the interactions favouring easy-axis states.
The equilibrium orientations of 
the staggered magnetization, 
derived from the N{\'e}el Eq.~(\ref{NeelEq}),
result from the competition between 
these interactions and the applied field.
Thus, the field $H_0$ 
sets the characteristic scale 
in the major part of the magnetic phase diagram.
Note, this is the only 
material parameter 
in the N\'eel equation (\ref{NeelEq}).

In the vicinity of the SF field, 
where two of the main energy contributions cancel each other, 
much weaker (relativistic) interactions enter 
the scene set by Eq.~(\ref{energy3}).
The  characteristic scale
in this region is given by the field
$H_{\textrm{c}}$, as introduced in Eq.~(\ref{Hc}).
Hence, $|H_{\textrm{c}}| \ll H_0$ defines a ``fine'' scale
of the system.
This scale gives the value of the potential barrier
$\Delta w_0 (H_{\textrm{SF}})= H_{\textrm{c}} M_0$
at the spin-flop field and it fixes the size of the 
metastable regions around the field $H_0$, 
i.e. the astroids.
The field $|H_{\textrm{c}}|$ includes two physically different 
contributions. 
One of them is the ratio between
the sublattice second-order anisotropy $K$ and
the antiferromagnetic coupling. 
The other is
the ratio between fourth- 
and second-order anisotropies, 
$B_2$ and $B_1$, of the staggered vector
(see Eq.~(\ref{uniaxial2})). 
Generally, the two terms have same order of magnitude. 

The strengths of the magnetic couplings in 
usual antiferromagnetic materials 
obey a well-defined hierarchy 
with very strong exchange and weaker uniaxial anisotropy.
This hierarchy is given 
by the relations $J \gg K, K', B_1 \gg B_2$.
Hence, the field  $H_0 = H_{\textrm{e}} \sqrt{B_1/(2J)}$
is much smaller than the exchange field $H_{\textrm{e}}$,
and $H_{\textrm{c}}$ = $(K/J+B_2/B_1)H_0/2$ is much smaller 
than $H_0$.
Correspondingly, the jump of the magnetization 
at the SF transition is small
$\Delta m \sim  H_0/H_{\textrm{e}} \ll 1$ (see Eqs.~(\ref{mP1}), (\ref{mP2})).
The potential barrier separating the stable states
at the SF field $\Delta w(H_{\textrm{SF}})= \varkappa B_1M_0^2$ 
is again much smaller than the barrier 
in the ground state, $\Delta w (0)= B_1M_0^2$,
and the region of the metastable states is restricted 
to a close vicinity of the SF field, $\Delta H \sim H_{\textrm{c}}$.
This causes the unusually high sensitivity 
of the magnetic states near SF field
with respect to small changes 
in strength and direction of applied fields.
For example, in an antiferromagnet with $ \varkappa > 0$ the
rotation of the magnetic field $H = H_{\textrm{SF}}$ from $\psi = -\psi_{\textrm{c}}$
to $\psi = \psi_{\textrm{c}}$ causes a change of the staggered
magnetization from $\theta = \pi/4$ to $\theta = -\pi/4$.
This lability of the magnetic states 
is the underlying reason for 
the remarkable magnetic effects in the SF region.

\section{Demagnetization effects and multidomain states}\label{domains2}
In the previous section  the equilibrium magnetic 
configurations have been derived as functions
of the internal magnetic field $\mathbf{H}$,
which differs from the applied
\textit{external} field $\mathbf{H}^{(\textrm{e})}$ 
due to the demagnetization field 
$\mathbf{H}^{(\textrm{m})}$ of the sample.\cite{Hubert98}
In a homogeneneously magnetized ellipsoidal sample
(including the limiting cases, i.e.
plates and long cylinders) the equation
\begin{eqnarray}
\mathbf{H}^{(\textrm{e})}=\mathbf{H}+
4\pi \widehat{\mathbf{N}}\mathbf{M}(\mathbf{H}), 
\label{He}
\end{eqnarray} 
establishes the relation between the external
and internal magnetic fields.
This relation allows to express the solutions 
for magnetic states
as functions of the external field by using
the demagnetization tensor $\widehat{\mathbf{N}}$ 
of the sample.
However, the relation between internal and external fields breaks
down at field-induced phase transitions.\cite{UFN88,Hubert98}
In the vicinity of such transitions the homogeneous states are
unstable with respect to the transformation into multidomain
structures consisting of domains formed from the competing phases.
\cite{UFN88,Hubert98}
Within the \textit{thermodynamic} approximation or  generalized
phase theory, two-phase multidomain states are described 
by the equation 
\begin{eqnarray}
\mathbf{H}^{(\textrm{e})}=\mathbf{H}_{tr}+4\pi \widehat{\mathbf{N}}<\mathbf{M}>, 
\label{energyD1}
\end{eqnarray}
where $\mathbf{H}_{tr}$ is the field value for the 
first-order transition.\cite{Hubert98,UFN88}
Here, $<\mathbf{M}>=\xi_1\mathbf{M}^{(1)}+\xi_2\mathbf{M}^{(2)}$,
$\mathbf{M}_i$ is the total average magnetization,
and the variable parameters $\xi_i$ 
are the volume fractions of the coexisting phases ($i=1, 2$,
$\xi_1 + \xi_2 = 1$).
For the SF transition the phase theory approximation
has been found to be valid practically in all regions
of the phase diagram,
where multidomain states exist.\cite{FTT82,UFN88}

\begin{figure*}
\includegraphics[width=7.5cm]{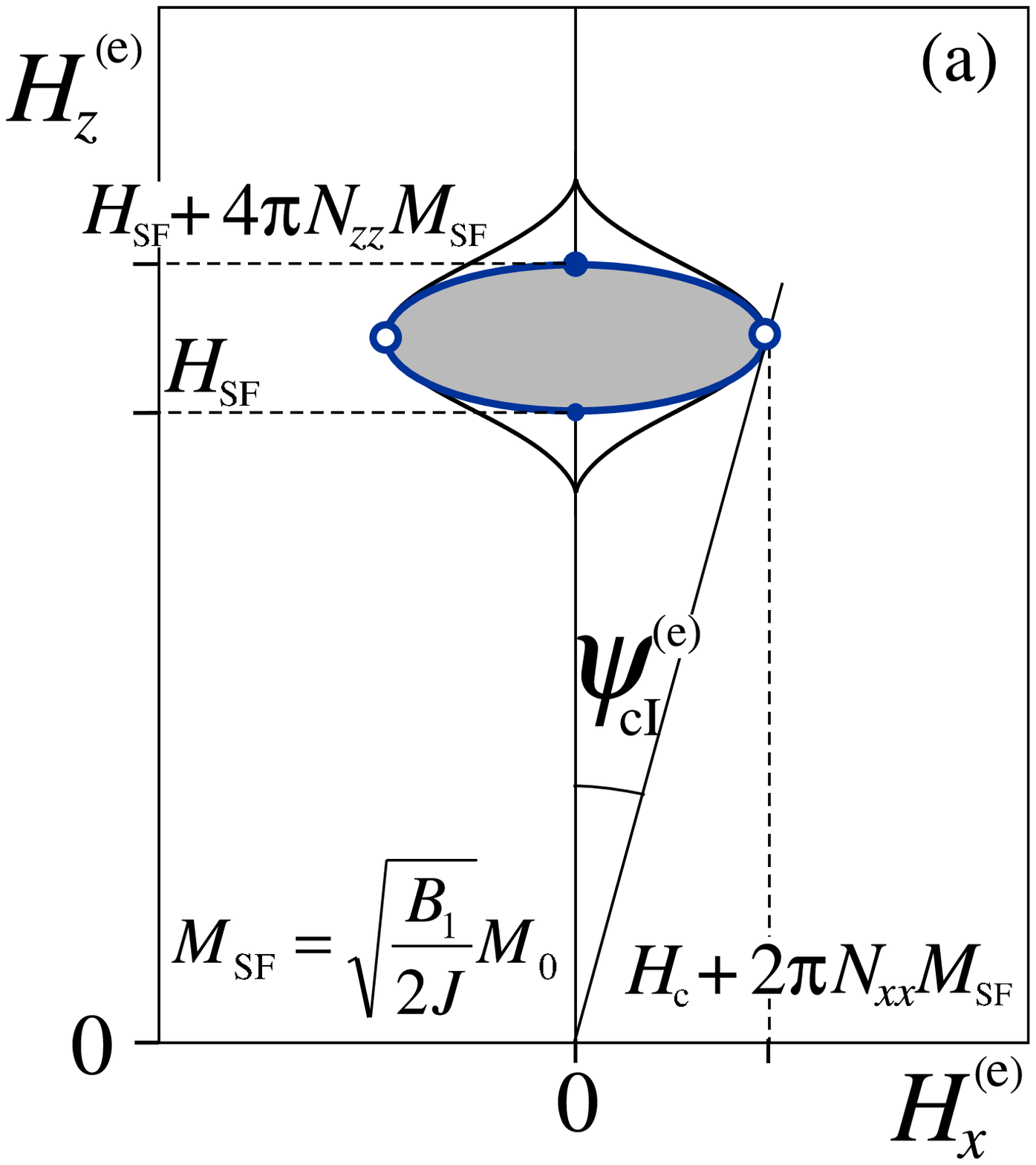}
\hspace{0.4cm}
\includegraphics[width=7.5cm]{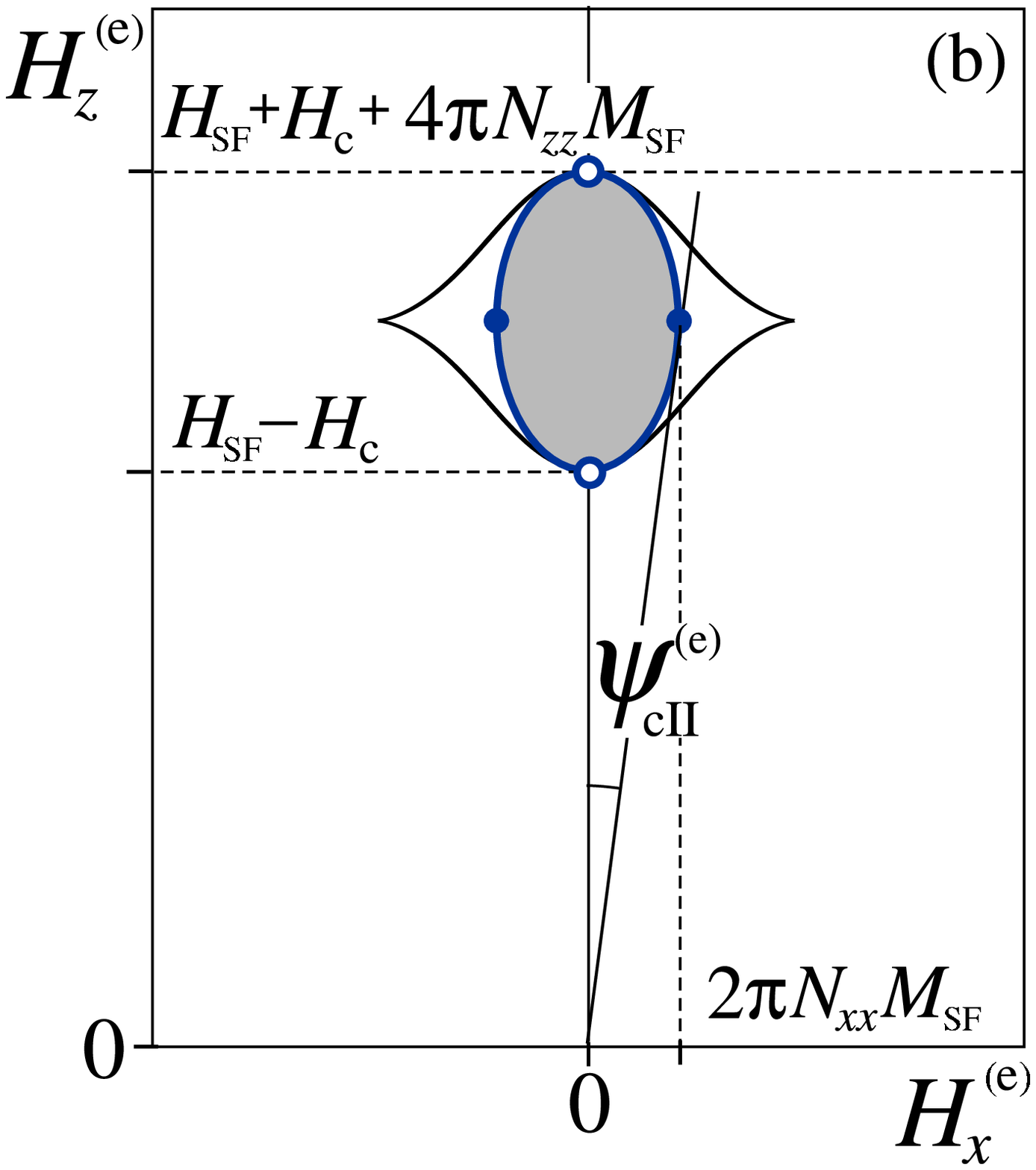}
\caption{
\label{DiagramExternal}
Magnetic phase diagrams in components of
the external field ($H_x^{(\textrm{e})}, H_z^{(\textrm{e})})$
for $\varkappa > 0$ (a) and $\varkappa < 0$ (b)
include regions of the multidomain states 
(shaded areas). 
}
\end{figure*}
For model (\ref{Phi4}) with
$\varkappa >0$
the phase theory equations (\ref{energyD1}) 
can be written in the following form
\begin{eqnarray}
H_x^{(\textrm{e})} & = & H_x \left(1 + \eta N_{xx} \right) 
\label{phaseAFM}\\
&  &+  \eta N_{xz} (\xi_1 - \xi_2) \sqrt{H_{\textrm{c}}^2 -H_x^2} 
+ \eta N_{xz} H_{\textrm{c}},
\nonumber\\
H_z^{(\textrm{e})} & = & H_{\textrm{SF}}+ H_x \tilde{\eta} N_{xz}  \nonumber\\
& &+ \eta N_{zz} (\xi_1 - \xi_2) \sqrt{H_{\textrm{c}}^2 -H_x^2}
+ \eta N_{zz} H_{\textrm{c}}\,,\nonumber
\end{eqnarray}
where $H_x$ varies along the first-order transition 
line ( $|H_x| \leq H_{\textrm{c}}$, $H_z = H_{\textrm{SF}}$).
The parameter $\eta = 2 \pi M_{\textrm{SF}}/H_{\textrm{c}} =\pi/(\varkappa J)$
measures the ratio of the stray field energy and
the potential barrier between the coexisting phases
at the SF transition.
Equations~(\ref{phaseAFM})
allow us to derive the parameters of
the multidomain states, $H_x$, $\xi_i$,
as functions of the external field.
In particular, for the relevant case with $N_{xz} =0$,  
Eqs.(\ref{phaseAFM}) with $\xi_{1(2)} =1$ yield 
the boundary of the multidomain states 
as an ellipse 
\begin{eqnarray}
1\!\!=\!\!
\left[\frac{H_{x}^{(\textrm{e})}}
{H_{\textrm{c}}+2 \pi N_{xx}M_{\textrm{SF}}}\right]^2 
\!\!\!+\!\!
\left[\frac{H_{z}^{(\textrm{e})}-H_{\textrm{SF}}-2 \pi N_{zz}M_{\textrm{SF}}}
{2 \pi N_{zz}M_{\textrm{SF}}}\right]^2  
\label{el1}
\end{eqnarray}
with semiaxes $a=H_{\textrm{c}}+2 \pi N_{xx}M_{\textrm{SF}}$, $b=2 \pi N_{zz}M_{\textrm{SF}}$
(Fig.~\ref{DiagramExternal}(A)).
For $\varkappa < 0$  similar equations 
yield the boundary ellipse
\begin{eqnarray}
1\!\!=\!\!
\left[\frac{H_{x}^{(\textrm{e})}}
{2\pi N_{xx}M_{\textrm{SF}}}\right]^2 
\!\!\!+\!\!
\left[\frac{H_z^{(\textrm{e})}-H_{\textrm{SF}}-2\pi N_{zz}M_{\textrm{SF}}}
{H_{\textrm{c}}+2\pi N_{zz}M_{\textrm{SF}}}\right]^2 
\label{el2}
\end{eqnarray}
with semiaxes $a= 2\pi N_{xx}M_{\textrm{SF}}$, 
$b = H_{\textrm{c}}+2\pi N_{zz}M_{\textrm{SF}}$
(Fig.~\ref{DiagramExternal}(B)).

The largest tilt angles between the applied
field and the easy axis, 
at which the multidomain states still exist,
can be readily derived from Eqs.~(\ref{el1}), (\ref{el2}),
%
%
%
\begin{eqnarray}
\psi_{\textrm{cI}}^{(\textrm{e})} &  = &  \frac{H_{\textrm{c}} + 2 \pi N_{xx} M_{\textrm{SF}}}{H_{\textrm{SF}}} 
= \psi_{\textrm{c}}  + \frac{\pi N_{xx}}{J}\textrm{\ for\ } \varkappa>0 \,,
\nonumber\\
\psi_{\textrm{cII}}^{(\textrm{e})} & = &  
\frac{2 \pi N_{xx} M_{\textrm{SF}}}{H_{\textrm{SF}}} = \frac{\pi N_{xx}}{J}
\textrm{\  for\ } \varkappa<0
\,.
\label{psimax}
\end{eqnarray}
The phase diagrams in Fig.~\ref{DiagramExternal} 
demonstrate the strong formal resemblance for
the two qualitatively different cases.
In both cases thermodynamically stable multidomain
states arise near the SF transition 
in a close vicinity to the SF field.
However, due to the different character of 
the phase transitions 
for $\varkappa > 0$ and $\varkappa < 0$, 
Eqs. (\ref{solution1}) and \ref{solution2}),
the evolution of the magnetic states 
within the multidomain regions (\ref{el1})
and (\ref{el2}) is different.
For $\varkappa > 0$ the competing phases
(\ref{solution1}) coexist along the horizontal line segment
($H_z = H_{\textrm{SF}}$, $|H_x|\leq H_{\textrm{c}}$).
In the ($H_x^{(\textrm{e})}, H_x^{(\textrm{e})}$) phase plane 
a set of the vertical straight lines describes 
regions with these fixed transition fields
(Fig. \ref{DiagramExternal} (a)).
Due to the smallness of $\psi_{\textrm{cI}}^{(\textrm{e})}$ (\ref{psimax}) 
external magnetic fields with fixed
directions ($|\psi^{(\textrm{e})}| \leq \psi_{\textrm{cI}}^{(\textrm{e})}$
practically intersect the region of the multidomain states
(\ref{el1}) along the lines with a fixed transition field.
A variation of fields with such a fixed orientation
causes in the system magnetization processes 
through the displacement of domain walls between
the coexisting states.
On the other hand, rotating magnetic 
fields with fixed amplitude that cross 
the region (\ref{el1})
mainly cause a continuous deformation of 
the magnetic configurations within the domains.
For  $\varkappa < 0$  rotating
fields cross the multidomain region
(\ref{el2}) almost perfectly along
lines corresponding to fixed internal fields.
Thus, a rotating field induces
a redistribution of the domains. 
On the other hand, magnetic fields with 
fixed directions produce mainly 
reorientation effects within the domains. 
\begin{figure*}[thb]
\includegraphics[width=8.75cm]{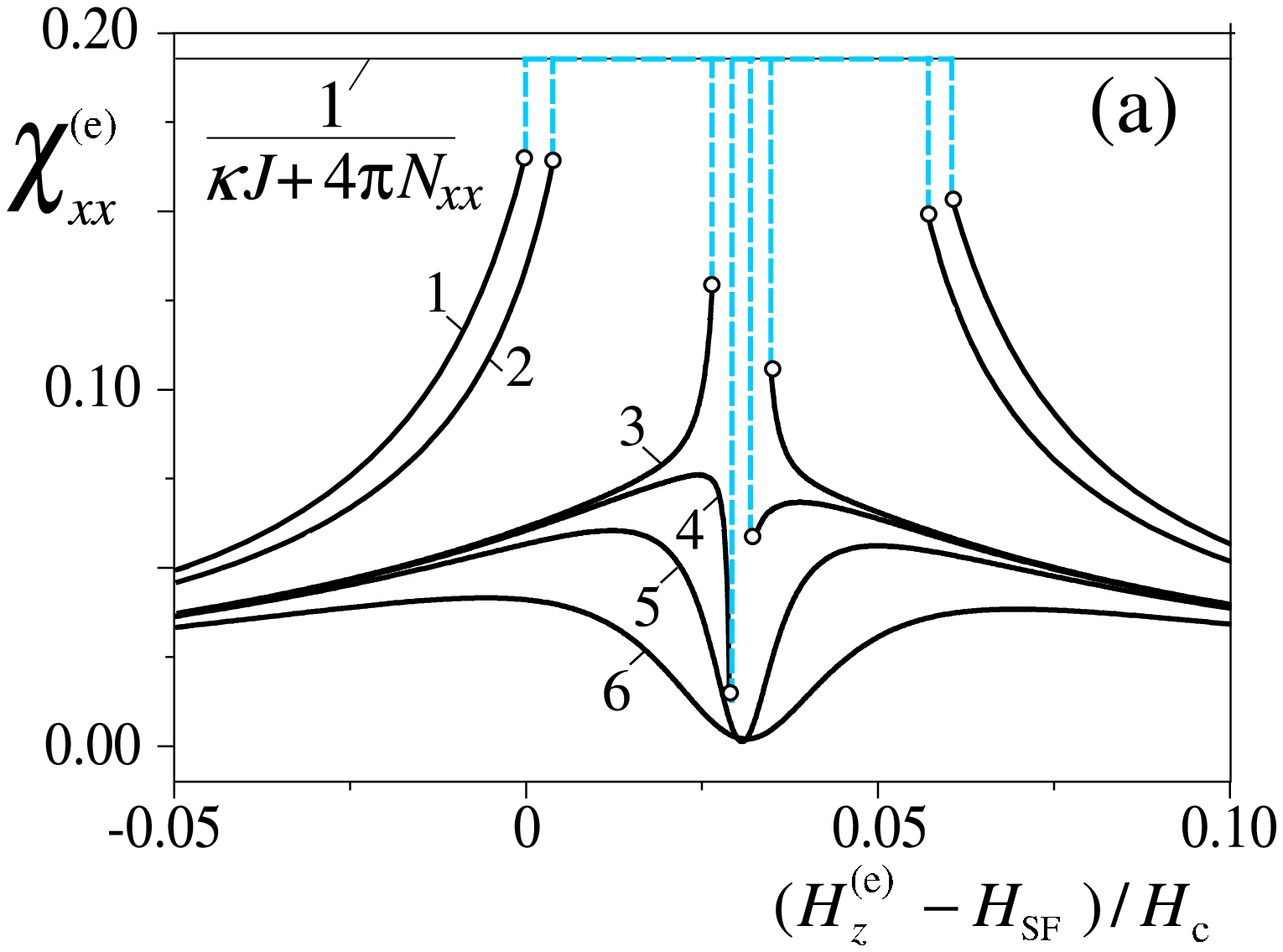}
\hspace{0.25cm}
\includegraphics[width=7.25cm]{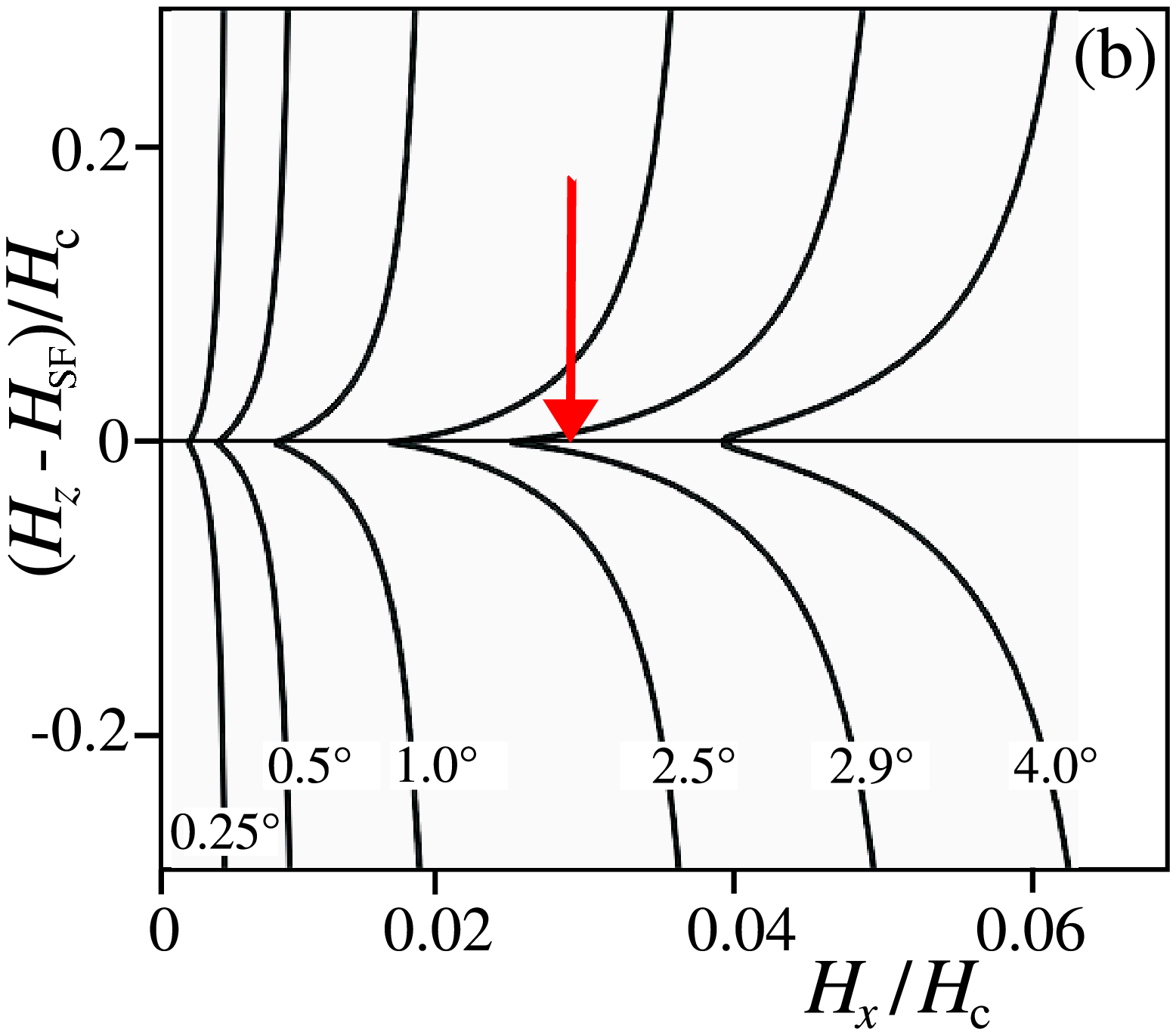}
\caption{
\label{ChixxTheory}
Dependencies of
$\chi_{xx}^{(\textrm{e})}$ on external field 
calculated for (C$_2$H$_5$NH$_3$)$_2$CuCl$_4$
for different fixed values of $\psi^{(\textrm{e})}$: 
$0.1^{\circ} (1), 
1.50^{\circ} (2),
2.86^{\circ} (3),
2.92^{\circ} (4),
2.98^{\circ} (5),
3.50^{\circ} (6)$.
Black solid lines correspond
to the homogeneous phases and dashed (blue)
lines to the multidomain states.
Hollow points indicate the boundaries 
between these regions.
Variations of the
internal magnetic field
$\mathbf{H} (H^{(\textrm{e})})$
for fixed directions of
the external field,
$\psi^{(\textrm{e})}$ = const (b).
The (red) arrow in (b) indicates
the location of the endpoint of
first-order transition.
The curved ``trajectories'' $H(H^{(\textrm{e})}$
explain the reentrant character of 
the $\chi_{xx}^{(\textrm{e})}(H^{(\textrm{e})})$
functions in (a).
}
\end{figure*}
\begin{figure}
\includegraphics[width=8.5cm]{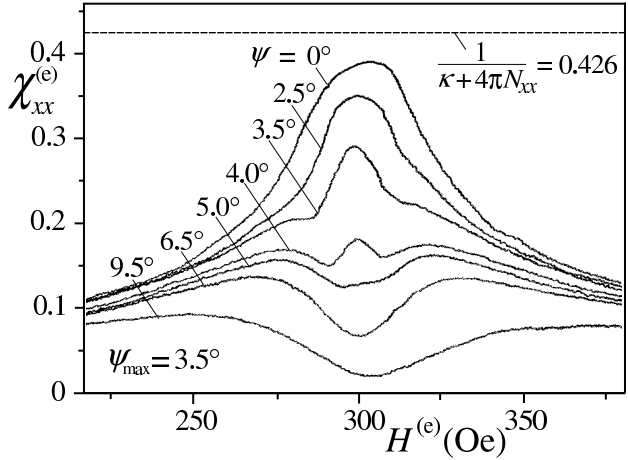}
\caption{
\label{ChixxExp}
Experimental dependencies of $\chi_{xx}^{(\textrm{e})}$
on the external field $H^{(\textrm{e})}$ for a number
of fixed angles $\psi^{(\textrm{e})}$  for 
a sphere of (C$_2$H$_5$NH$_3$)$_2$ CuCl$_4$ 
(Sample No.~1) at T = 4.2 K. 
 }
\end{figure}

Correspondingly, the limiting angles (\ref{psimax})
have a different physical meaning in both cases.
For systems with $\varkappa > 0$ the critical points
at $\psi^{(\textrm{e})} = \pm \psi_{\textrm{cI}}^{(\textrm{e})}$,
$H^{(\textrm{e})} = H_{\textrm{SF}} + 2 \pi N_{zz} M_{\textrm{SF}}$
correspond to internal states 
where the difference between 
the magnetic configurations 
in the coexisting phases disappears.
For antiferromagnets with $\varkappa < 0$ 
the transition into the homogeneous state at
the critical points 
$\psi^{(\textrm{e})} = \pm \psi_{\textrm{cII}}^{(\textrm{e})}$,
$H^{(\textrm{e})} = H_{\textrm{SF}} + 2 \pi N_{zz} M_{\textrm{SF}}$
occurs by a complete replacement of one of 
the coexisting phases by the other through domain processes.

The quantitative description of the magnetization
in the multidomain states is provided by
Eq.~(\ref{energyD1}).
This equation together with those for $\mathbf{M} (\mathbf{H})$ 
(\ref{mP1}), (\ref{mP2}) yields  the functions
$\langle \mathbf{M} \rangle (\mathbf{H}^{(\textrm{e})})$.
It follows from (\ref{energyD1})
that the magnetization is a linear functions
of the applied field only when 
$\mathbf{H}^{(\textrm{e})}$ varies along
lines corresponding to fixed internal transition fields.
In the general case the variation of the transition
field $\mathbf{H}^{(\textrm{t})} (\mathbf{H}^{(\textrm{e})})$
causes complex dependencies of the magnetization 
on the external field in the multidomain state.
Differentiation of $\langle \mathbf{M} \rangle$ 
in Eq.~(\ref{energyD1})
yields the equation for the components of
the magnetic susceptibility (cf.~$[$\onlinecite{UFN88}$]$)
\begin{eqnarray}
4 \pi  {N}_{ij} \chi_{jk}= \delta_{ik} 
- \frac{\partial H_{i}^{(\textrm{t})}}{\partial H_k^{(\textrm{e})}}\,.
\label{chiD}
\end{eqnarray}
The first term on the right side of (\ref{chiD}) 
describes the process of a 
redistribution of volume fractions for the different 
phases  through \textit{displacement of domain walls},
while the second term is associated with the variation
of the magnetic states within the domains by \textit{rotations
of the magnetization}.
In homogeneous phases 
Eq.~(\ref{He}) allows one to 
relate the external susceptibility $\hat{\chi}^{(\textrm{e})}$ and 
the internal susceptibility $\hat{\chi}$
\begin{eqnarray}
\hat{\chi}^{(\textrm{e})} = \left( \hat{\chi}^{-1}
+ 4 \pi  \hat{N} \right)^{-1}\,.
\label{chiEx}
\end{eqnarray}
Eqs.~(\ref{energyD1}), (\ref{chiD}), (\ref{chiEx})
transform the components of the net magnetization
(\ref{m3}) and magnetic susceptibility
(\ref{chixx}), (\ref{chixz}), (\ref{chizz}) derived
in the components of the internal field into
those in components of the external field.
The tensor of external susceptibility (\ref{chiEx})
includes  the internal susceptibility  $\hat{\chi}$
and the stray field contribution  
$\hat{\chi}_{m} =(4 \pi  \hat{N})^{-1}$ known
as \textit{shape susceptibility}.\cite{UFN88}
In the multidomain state with fields varying along 
the lines of the fixed transition field 
the external susceptibility as given by  (\ref{chiD}) 
has only contributions from the shape susceptibility,
$4 \pi  {N}_{ij} \chi_{jk}= \delta_{ik} $.
When the evolution of a multidomain state
involves a variation of the transition field
a specific susceptibility contribution arises
that is associated with the rotation of 
the magnetic states within in the domains.
In particular, for the multidomain
states in Fig.~\ref{DiagramExternal} 
along the line $H_x^{(\textrm{e})} =0$ 
\begin{eqnarray}
\chi_{zz}^{(\textrm{e})} & =& \frac{1}{4\pi N_{zz}} \quad (\varkappa > 0),
\label{chiEA}\\
\chi_{zz}^{(\textrm{e})} & = & \frac{1}{4 |\varkappa| J+ 4\pi N_{zz}} 
\quad (\varkappa < 0)\,. \nonumber
\end{eqnarray}
For $\varkappa > 0$ the magnetic field varies along the line of the
fixed transition field, and the susceptibility includes 
only shape contribution.
For $\varkappa < 0$ the magnetic field varies along the line
with fixed volume fractions, $ \xi_1=\xi_2= 1/2$, and the evolution
of the system consists of a continuous reorientation  
in the domains (\ref{solution2}). 
For this process the net
magnetization is derived from Eq.~(\ref{mP2}), and the
internal susceptibility is $\chi_{zz} = 1/(4 |\varkappa| J)$.
The external susceptibility $\chi_{zz}^{(\textrm{e})}$ (\ref{chiEA})
includes both internal and shape contributions.
Generally during the SF transition the values of 
the internal susceptibility  
$~ 1/(|\varkappa| J)$ (\ref{chixx}), (\ref{chixz}), (\ref{chizz})
arising due to variation of the homogeneous magnetic states
and the shape susceptiblity $~ 1/N$ originating from the 
reconstruction of the multidomain states are of 
the same order and larger  than the values for 
the external  susceptibility outside of the SF region. 
For example, in the external field 
along the easy axis  $\chi^{(\textrm{e})}$
equals zero in the AF phase, and in the SF phase 
$\chi_{zz}^{(\textrm{e})}=1/(2J + 4\pi N_{zz})$.

Fig.~\ref{ChixxTheory} (a) shows the calculated
external-field dependencies of $\chi_{xx}(H^{(\textrm{e})})$ 
for a number of fixed angles $\psi^{(\textrm{e})}$ and with materials
parameters close to those for 
(C$_2$H$_5$NH$_3$)$_2$CuCl$_4$.
According to previous investigations 
in this antiferromagnet,
$\varkappa > 0$.\cite{FNT89,JMMM05}
The functions $\chi_{xx}(H^{(\textrm{e})})$ 
for small angles $\psi^{(\textrm{e})} < \psi_{\textrm{cI}}^{(\textrm{e})}$, 
are given by \textit{lines 1-3} in Fig.~\ref{ChixxTheory} (a).
They cross field-dependent regions
corresponding to homogeneous phases (black solid lines)
and areas with constant susceptibilities (dashed blue lines)
indicating multidomain regions.
For $\psi^{(\textrm{e})} < \psi_{\textrm{cI}}^{(\textrm{e})}$ the functions 
$\chi_{xx}(H^{(\textrm{e})})$
(\textit{lines 5, 6}) have characteristic 
features as described in the previous
section (Figs.~\ref{Fchixx} 
and Eqs.~(\ref{MaxXX1}), (\ref{MaxXX2})).
The intermediate \textit{line 4} shows ambivalent features.
In the center it includes a jump into 
the multidomain states as  the lines
for $\psi^{(\textrm{e})} < \psi_{\textrm{cI}}^{(\textrm{e})}$,
but outside this narrow central feature the function 
$\chi_{xx}(H^{(\textrm{e})}$ behaves similar to those 
for $\psi^{(\textrm{e})} >\psi_{\textrm{cI}}^{(\textrm{e})}$.
This interesting effect is explained 
by strong deviations of the total magnetization 
from the easy axis in the vicinity of SF field.
The enhanced values of the tranverse magnetization
$M_x$ (Fig.~\ref{magnetizationSF} (c)) create
a strong demagnetization screening near the SF field.
Accordingly, for the external field  varying along
lines $\psi^{(\textrm{e})} =const$, interal field ``trajectories''
$H(H^{(\textrm{e})})$ deviate towards the $H_z$ axis near the SF field,
as depicted in Fig.~\ref{ChixxTheory} (b).
The trajectory for the evolution of the internal field $\mathbf{H}$ 
starts from outside the transition region
i.e. $|H_x| > H_{\textrm{c}}$ but enters this area 
in the vicinity of the SF field
for the external field applied 
under an angle $\psi^{(\textrm{e})} =2^{\circ}$.

\section{Magnetic phase diagram 
of (C$_2$H$_5$NH$_3$)$_2$CuCl$_4$ }\label{experiment}
\begin{figure}
\includegraphics[width=8.5cm]{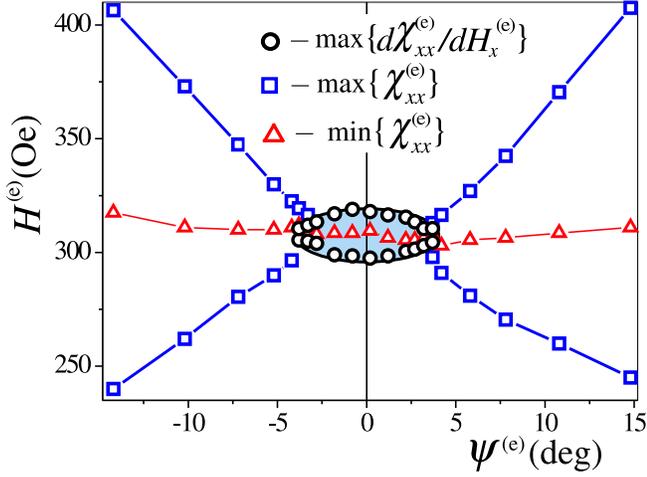}
\caption{
\label{ChixxExp2}
Location of extremal points of  $\chi_{xx}^{(\textrm{e})}$ 
and $\partial \chi_{xx}^{(\textrm{e})}/\partial H^{(\textrm{e})}$
in the phase plane ($H^{(\textrm{e})}, \psi^{(\textrm{e})}$)
for an elliptical (C$_2$H$_5$NH$_3$)$_2$ CuCl$_4$ sample
(No. 3).
The shaded area indicates the region of multidomain
states.
}
\end{figure}
\begin{figure}
\includegraphics[width=8.5cm]{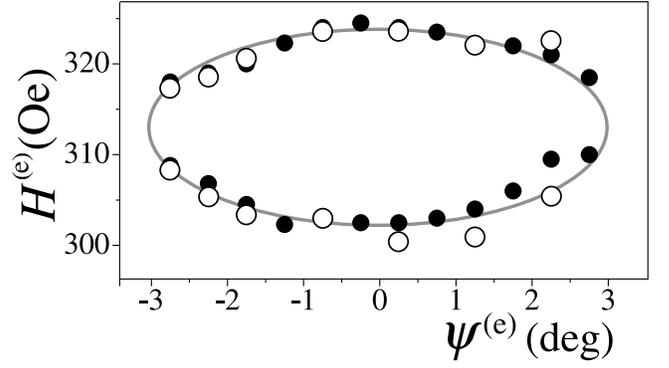}
\caption{
\label{domains}
Experimental phase diagram in components
of applied field for 
(C$_2$H$_5$NH$_3$)$_2$ CuCl$_4$
(for spherical and ellipsoidal samples). 
The region of multidomain states
is within the area circumscribed by measured
points. Filled circles give the field values 
for the maxima 
in $\partial \chi_{xz}/\partial H_z^{(\textrm{e})} ( H_z^{(\textrm{e})})$;
hollow circles give the inflection points in
$\partial \chi_{zz}/\partial H_z^{(\textrm{e})} ( H_z^{(\textrm{e})})$.
}
\end{figure}

An experimental investigations of the differential magnetic
susceptibility was carried out for 
(C$_2$H$_5$NH$_3$)$_2$CuCl$_4$.\cite{ZhurNEXT}
This model antiferromagnet 
has orthorhombic lattice structure with space group is Pbca, 
and orders at a N\'eel temperature T$_N$ = 10.20 K.\cite{Jongh72}.
The magnetic structure consists of ferromagnetic layers
parallel to the $x0z$ planes in the notation of this paper
with weak antiferromagnetic couplings.
According to Ref.~$[$\onlinecite{Jongh72}$]$ 
the ferromagnetic intra-layer
interactions corresponds to an effective field
500~kOe, while antiferromagnetic coupling between layers
is $JM_0$ = 837.5 Oe. 
A weak second-order 
anisotropy in the $x0z$ plane $KM_0$ = 76 Oe
stabilizes the collinear ground state
with the sublattice magnetizations $\mathbf{M}_i$
along the $z$-axis.
At T = 4.2 K  the spin-flop field is at
$H_{\textrm{SF}}$ = 305 Oe.
Magnetic interactions in 
(C$_2$H$_5$NH$_3$)$_2$CuCl$_4$
include a Dzyaloshinsky-Moriya coupling
described by an energy contribution
$w_D = D(M_{1x}M_{2y}-M_{2x}M_{1y})$
with an effective field $DM_0$= 119 Oe.
\cite{Jongh72}
Generally this interaction  
rotates the magnetization vectors $\mathbf{M}_i$
away from the basal plane $x0z$. 
However, due to the strong orthorhombic anisotropy
with a value of 1504 Oe, the vectors $\mathbf{M}_i$ are 
practically confined to the basal plane.
Deviations from this plane 
do not exceed 2$^{\circ}$.\cite{FNT89}
Thus, the  planar model (\ref{energy3}) 
can be applied to this antiferromagnet.

The differential magnetic susceptibility
components were measured with an inductive 
technique using three pairs of modulating
and pick-up coils arranged along perpendicular
directions. The modulating fields had
amplitudes between 0.3 and 1.0 Oe at
frequencies of 9, 17-21, 86 and 133 Hz.
The three samples used for the measurements 
were cut from a (C$_2$H$_5$NH$_3$)$_2$CuCl$_4$
single crystal. 
The samples had the following
geometrical parameters:
(Sample No. 1) a sphere with diameter 3.1 mm 
($N_{xx} = N_{zz} = 1/3$);
(Sample No. 2) an ellipsoid with
axes 
$a_x$ = 5.00 mm, $a_y$ = 1.75 mm,
$a_z$ = 5.00 mm
($N_{xx}$ =0.185,  $N_{zz}$ = 0.185);
(Sample No. 3) an ellipsoid with
$a_x$ = 3.43 mm, $a_y$ = 2.45 mm,
$a_z$ = 5.75 mm
($N_{xx}$ =0.344,  $N_{zz}$ = 0.186).
All the measurements of components
$\chi_{ij}^{(\textrm{e})}$ have been carried
out at $T=$~4.2~K. The recorded data
for all investigated samples
are in close accordance with the theoretical
results expounded in previous sections.
An example is shown in Fig.~(\ref{ChixxExp}),
where $\chi_{xx}^{(\textrm{e})}$ components are
plotted for the spherical sample.
The experimental data follow closely the theoretical results
of Eq.~(\ref{chixx}) as sketched in Fig.~\ref{ChixxTheory}.
In particular, the dependencies of $\chi_{xx}^{(\textrm{e})}(H^{(\textrm{e})})$ 
for $\psi^{(\textrm{e})} = 3.5^{\circ}, 4.0^{\circ}$ and $5^{\circ}$
display the reentrant behaviour imposed by 
the rotation of the internal field towards
the $H_z$-axis, as demonstrated in Fig.~\ref{ChixxTheory}.
By fitting experimental data for the $\chi_{ij}^{(\textrm{e})}$
components and their field derivatives 
with the theoretical dependencies,
values of the material parameters
can be deduced and the magnetic phase diagram of
this antiferromagnet has been constructed.
Fig.~\ref{ChixxExp2} shows the locations of
the extremal points for $\chi_{xx}^{(\textrm{e})}$
and $d \chi_{xx}^{(\textrm{e})}/d H^{(\textrm{e})}$
and the region of the multidomain states
in the ($H^{(\textrm{e})}, \psi^{(\textrm{e})}$) phase diagram
of the ellipsoidal samples.

The experimental ($H^{(\textrm{e})}, \psi^{(\textrm{e})}$)
phase diagram for the spherical sample
is shown in Fig.~\ref{domains}.
In particular, for the maximal angle
$\psi_{\textrm{cI}}^{(\textrm{e})}$ the following
results have been obtained
(spherical sample No.~1) $3.50^{\circ} \pm 0.65^{\circ}$ 
%
(sample No. 2) $2.70^{\circ} \pm 0.45^{\circ}$ 
%
(sample No. 3) $3.55^{\circ} \pm 0.65^{\circ}$ 
%
These results yield the values
of the spin-flop field $H_{\textrm{SF}} = 306$ Oe,
the critical angle
$\psi_{\textrm{c}} = 1.7^{\circ}$,
and
the characteristic fields
$H_0$ = 300 Oe and $H_{\textrm{c}}$ = 9.1 Oe.

\section{Conclusions}\label{conclusions}

Magnetic configurations for an extended class of
two-sublattice collinear easy-axis antiferromagnets 
have been obtained as functions of the values and 
directions of the applied magnetic field, 
and the corresponding magnetic phase
diagrams have been constructed.
The magnetic behaviour of these 
materials strongly depends on the
strengths of the applied field.
In the main part of the magnetic phase diagram
they are described by the
well-known N\'eel model (\ref{energy01}).
In the vicinity of the SF field they are described
by an effective model for a reorientation transition
that is formally equivalent to the Stoner-Wohlfarth model
of a uniaxial ferromagnet (\ref{Phi4}).
This \textit{two-scale} character
of easy-axis antiferromagnets has
been ignored in most previous investigations.
The analysis of magnetic-field-driven 
reorientation effects and the concomitant multidomain
states provides a consistent
picture of the magnetization 
processes near the SF transition. 
At the SF transition the weak intrinsic 
higher-order couplings of the antiferromagnetic 
material and the dipolar stray field 
cause important and noticeable effects
and must be included in the analysis.
These  effects are responsible for the
prominent anomalies in the magnetization 
curves and the magnetic susceptibilities
of antiferromagnets in the spin-flop region.
The anomalies provide effective approaches 
for investigations on magnetic properties of antiferromagnets 
that are usually hidden away from the spin-flop region.
The validity of this approach has been
demonstrated by an application to
an orthorhombic antiferromagnet.
 
The results on bulk antiferromagnets may also
be extended to confined antiferromagnetic
systems by including surface/interface-induced interactions 
into the phenomenological models.  
In Ref.~$[$\onlinecite{PRB03}$]$ 
it was shown that
the interplay between surface-induced and
intrinsic magnetic interactions yield
a rich variety of specific magnetic states
including spatially inhomogeneous twisted
states in the vicinity of the SF field.
The further development the theory for 
such cases will be important
for an understanding of the magnetization processes
in ferro/antiferromagnetic bilayers,\cite{Felcher96,Nogues00,bias}
and in antiferromagnetic nanoparticles.\cite{Hansen00}

The phenomenological model of the two-sublattice antiferromagnet
(\ref{energy0}) and its variants can be adopted also 
to describe magnetic states 
in synthetic antiferromagnets.\cite{Worledge04,APLNew}

\begin{acknowledgments}
This work was financially supported by DFG
through SPP 1133, project RO 2238/6-1.
A.\ N.\ B.\ thanks H.\ Eschrig for support and
hospitality at the IFW Dresden.
\end{acknowledgments}


\end{document}